\documentclass[runningheads,a4paper]{llncs}

\usepackage{amssymb}
\usepackage{graphicx}
\usepackage{url}
\usepackage{amsmath}
\usepackage{booktabs}

\usepackage{enumerate}
\usepackage{xcolor}
\usepackage{color}
\definecolor{darkgreen}{rgb}{0,0.6,0}
\definecolor{darkblue}{rgb}{0,0,0.6}

\usepackage{url}
\urldef{\mailsa}\path|{alfred.hofmann, ursula.barth, ingrid.haas, frank.holzwarth,|
\urldef{\mailsb}\path|anna.kramer, leonie.kunz, christine.reiss, nicole.sator,|
\urldef{\mailsc}\path|erika.siebert-cole, peter.strasser, lncs}@springer.com|    
\newcommand{\keywords}[1]{\par\addvspace\baselineskip
\noindent\keywordname\enspace\ignorespaces#1}

\begin{document}

\mainmatter  % start of an individual contribution

% first the title is needed
\title{On clustering financial time series: a need for distances between dependent random variables
}

% a short form should be given in case it is too long for the running head
\titlerunning{On clustering financial time series}

% the name(s) of the author(s) follow(s) next
%
% NB: Chinese authors should write their first names(s) in front of
% their surnames. This ensures that the names appear correctly in
% the running heads and the author index.
%
\author{Gautier Marti$^{1,2}$
\and Frank Nielsen$^{2}$ \and Philippe Donnat$^{1}$ \and S\'ebastien Andler$^{3}$}
\authorrunning{Lecture Notes in Computer Science: On clustering financial time series}
% (feature abused for this document to repeat the title also on left hand pages)

% the affiliations are given next; don't give your e-mail address
% unless you accept that it will be published
\institute{$^{1}$Hellebore Capital Management\\
63 Avenue des Champs-Elys\'ees, 75008 Paris, France\\
\vspace{0.5em}
$^{2}$Ecole Polytechnique\\
LIX - UMR 7161, 91128 Palaiseau Cedex, France\\
\vspace{0.5em}
$^{3}$Ecole normale sup\'erieure de Lyon\\
46 all\'ee d'Italie, 69364 Lyon Cedex 07, France
}

\toctitle{Lecture Notes in Computer Science}
\tocauthor{Authors' Instructions}
\maketitle

\begin{abstract}

The following working document summarizes our work on the clustering of financial time series.
It was written for a workshop on information geometry and its application for image and signal processing.
This workshop brought several experts in pure and applied mathematics together with applied researchers from medical imaging, radar signal processing and finance.
The authors belong to the latter group. This document was written as a long introduction to further development of geometric tools in financial applications such as risk or portfolio analysis. Indeed, risk and portfolio analysis essentially rely on covariance matrices. Besides that the Gaussian assumption is known to be inaccurate, covariance matrices are difficult to estimate from empirical data. To filter noise from the empirical estimate, Mantegna proposed using hierarchical clustering.
In this work, we first show that this procedure is statistically consistent.
Then, we propose to use clustering with a much broader application than the filtering of empirical covariance matrices from the estimate correlation coefficients. To be able to do that, we need to obtain distances between the financial time series that incorporate all the available information in these cross-dependent random processes.

\keywords{clustering; financial time series; noisy covariance matrix; dependence structure; distance between distributions; empirical finance; credit default swap}
\end{abstract}

\section{Clustering for financial risk modelling}

In financial applications, the variance-covariance matrix is an essential tool to assess the risk of a portfolio. Assuming that assets' returns are following a Gaussian multivariate distribution, the variance-covariance matrix captures both their joint behaviour (in this case, their Pearson correlation) and the specific risk of each asset which corresponds to its returns' standard deviation (also named volatility in finance). However, using an empirical variance-covariance matrix suffers from at least two shortcomings:
\begin{enumerate}[(i)]
\item if the assets' returns are following another multivariate distribution, then the variance-covariance matrix only measures a mixed information of linear dependence perturbed by the (possibly heavy-tailed) marginals. In this case, the variance-covariance matrix is not a relevant tool to quantify the risk between financial assets from their past returns time series; 
\item estimating the empirical variance-covariance matrix from data is a problem in itself \cite{laloux2000random}. For $N$ assets, one has to estimate $N(N-1)/2$ coefficients from $N$ time series of length $T$. If $T$ is small compared to $N$, the coefficients will be noisy and the matrix to some extent random.
\end{enumerate}

\begin{figure}
\begin{center}
\includegraphics[scale=0.5]{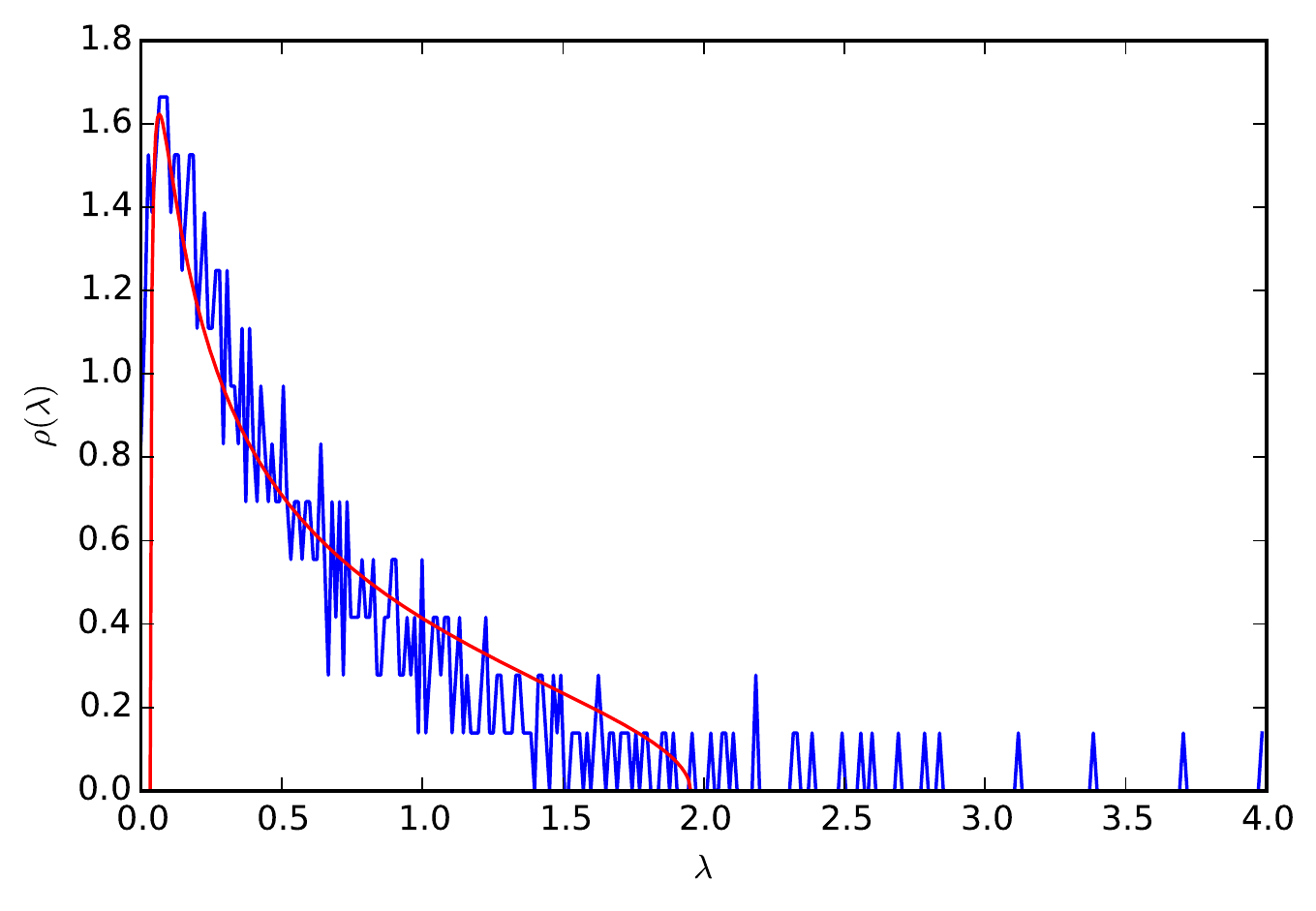}
\caption{Theoretical eigenvalues density for a purely random correlation matrix (red) vs. empirical density (blue) of the correlation matrix eigenvalues}\label{fig:mp_distrib}
\end{center}
\end{figure}

Shortcoming (ii) has been adressed in the literature by several approaches. One of them leverages results from the Random Matrix Theory (RMT) and can be found under the terms ``noise dressing" in the econophysics literature  \cite{laloux1999noise,laloux2000random,plerou2002random,potters2005financial,allez2014eigenvectors,bun2015rotational}.
For example, authors in \cite{laloux1999noise} compare the distribution of the empirical correlation eigenvalues to the known theoretical distribution given by RMT, and find that 94\% of the total number of eigenvalues falls in the support of the theoretical distribution. This experiment was led on stock market data, more precisely using $N = 406$ assets of the S\&P500 during the years 1991-1996. We can observe that this stylized fact about correlation between stocks also applies to different markets and different periods. For example, we illustrate this empirical property on the credit default swaps (CDS) market.
Let $X$ be the matrix storing the standardized daily returns of $N=560$ credit default swaps (5-year maturity) during the years 2006-2015 ($T \approx 2500$ values for each time series).
Then, the empirical correlation matrix of the returns is
$C = \frac{1}{T} X X^\top.$
We can compute the empirical density of its eigenvalues
$\rho(\lambda) = \frac{1}{N} \frac{dn(\lambda)}{d\lambda},$
where $n(\lambda)$ counts the number of eigenvalues of $C$ less than $\lambda$.
From random matrix theory, the limit distribution as $N \rightarrow \infty$, $T \rightarrow \infty$ and $T/N$ fixed reads:
\begin{equation}
\rho(\lambda) = \frac{T/N}{2\pi} \frac{\sqrt{(\lambda_{\max} - \lambda)(\lambda - \lambda_{\min})}}{\lambda},
\end{equation}
where $\lambda_{\min}^{\max} = 1 + N/T \pm 2\sqrt{N/T}$, and $\lambda \in [\lambda_{\min},\lambda_{\max}]$.
We can observe in Figure~\ref{fig:mp_distrib} that the theoretical distribution fits well the empirical one meaning that most of the information contained in the empirical correlation matrix can be considered noise. Only $26$ eigenvalues are greater than $\lambda_{\max}$, i.e. 95\% of the total number of eigenvalues falls in the support of the theoretical distribution.

\begin{figure}
\begin{center}
\includegraphics[scale=0.5]{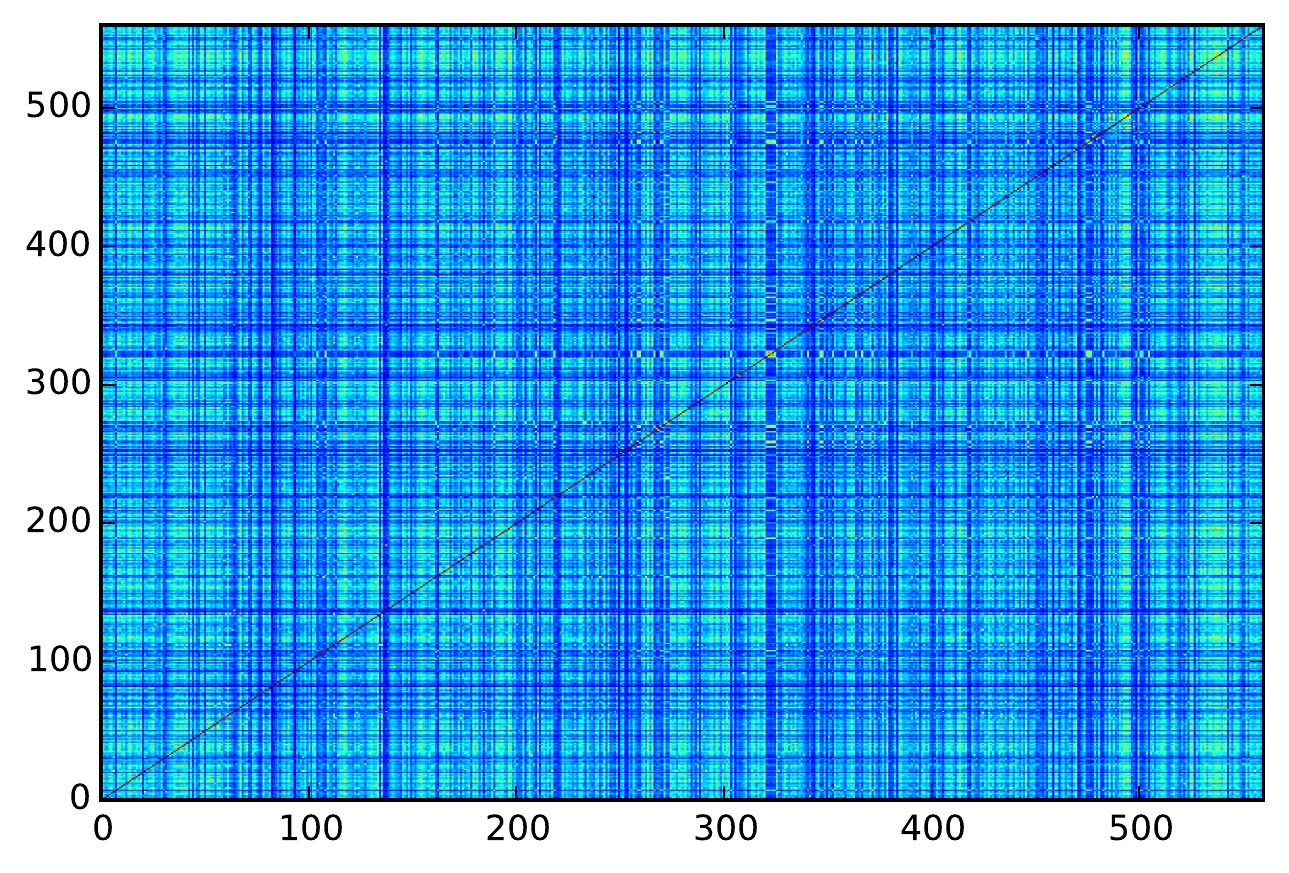}
\caption{An empirical and noisy correlation matrix computed on the log-returns of $N=560$ credit default swap time series of length $T\approx2500$}\label{fig:emp_cor}
\end{center}
\end{figure}

\begin{figure}
\begin{center}
\includegraphics[scale=0.47]{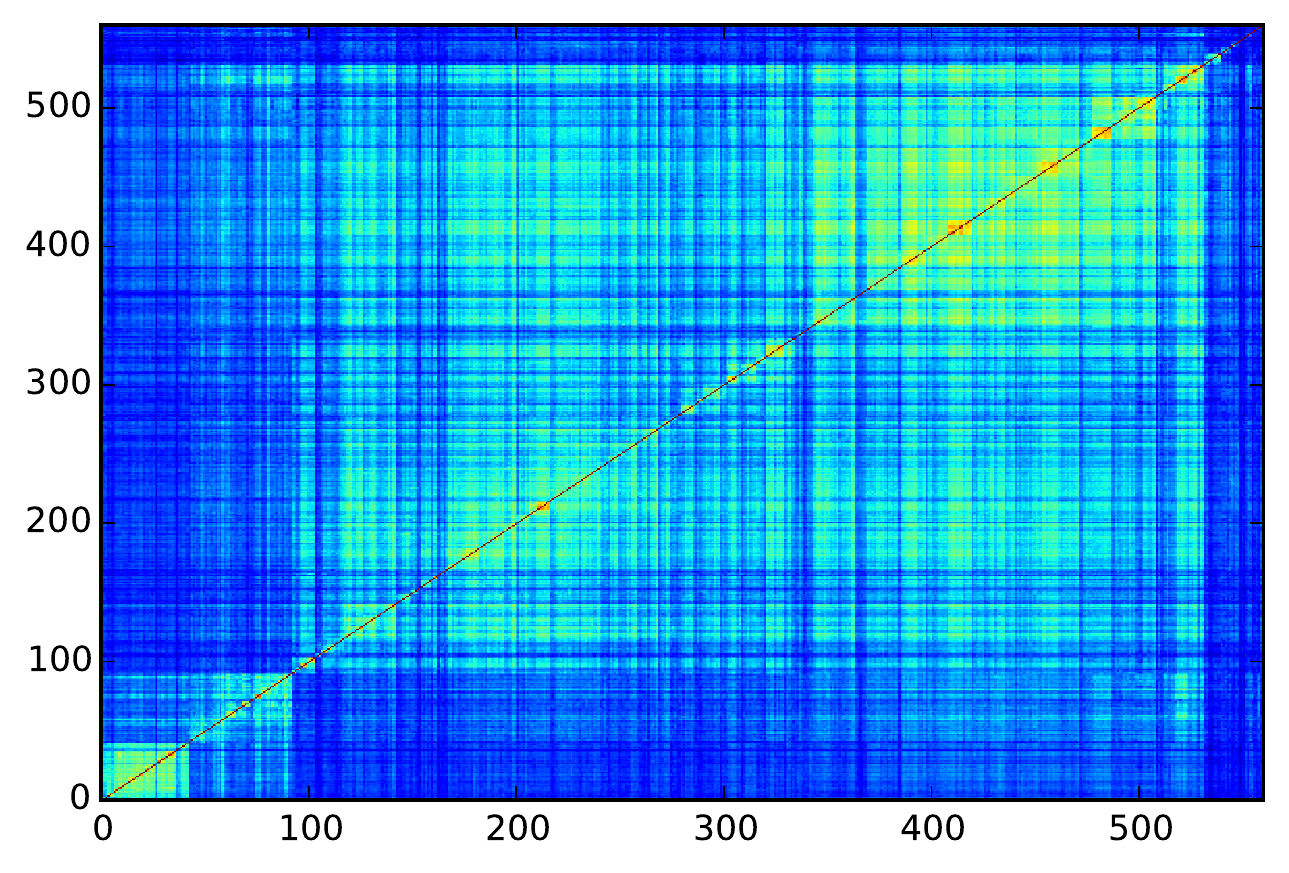}\includegraphics[scale=0.47]{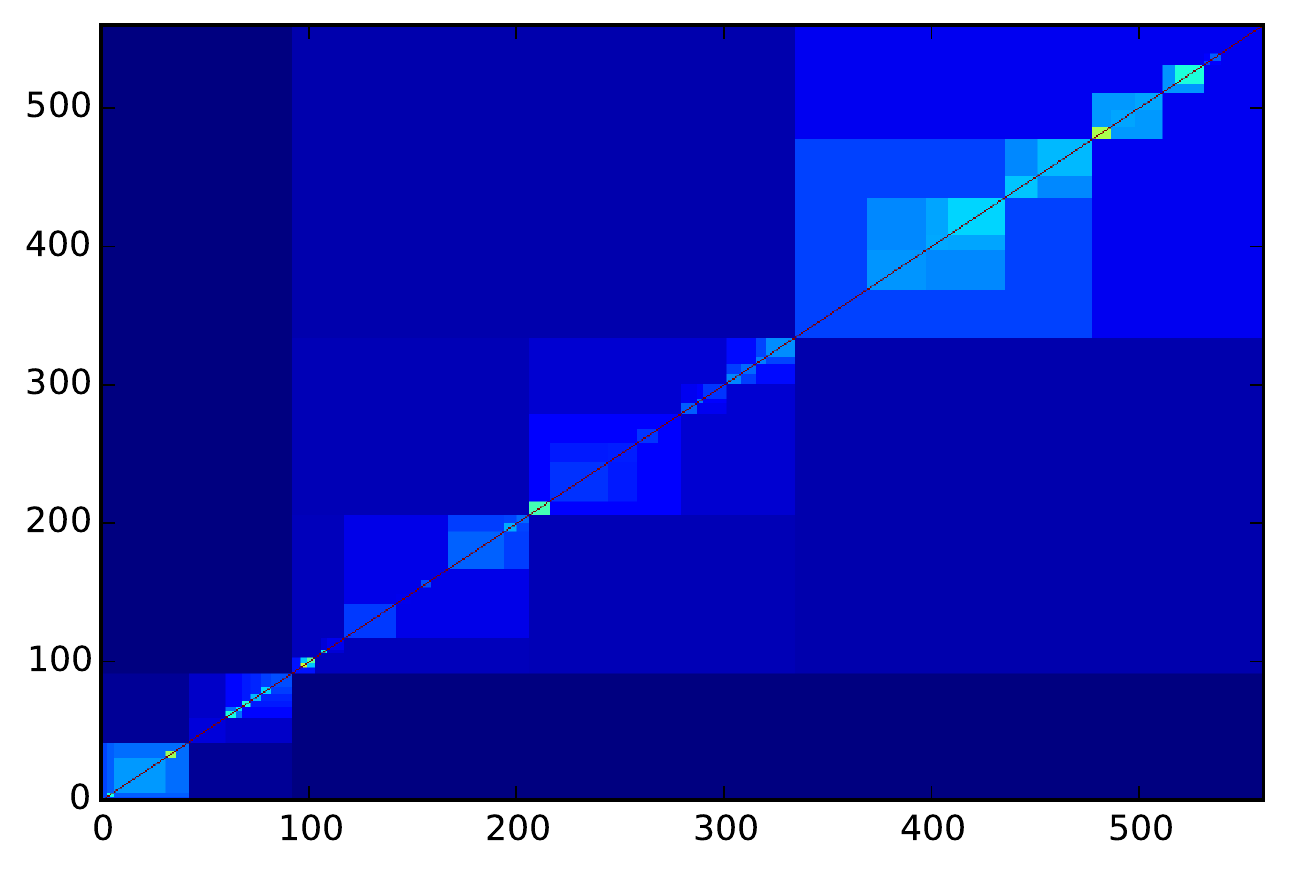}
\caption{The same noisy correlation matrix re-ordered by a hierarchical clustering algorithm; one can notice its noisy hierarchical correlation structure (left); The filtered correlation matrix resulting from the method described in \cite{mantegna1999introduction} (right) }\label{fig:emp_seriated}
\end{center}
\end{figure}

These results are important to take into account: for example, they have ``interesting
potential applications to risk management and
portfolio optimisation. It is clear [\ldots]
that Markowitz's portfolio optimisation scheme based on
a purely historical determination of the correlation matrix
is not adequate, since its lowest eigenvalues (corresponding
to the smallest risk portfolios) are dominated
by noise" \cite{laloux2000random}.
It motivates the need for filtering procedures of correlation matrices. Besides the RMT approach, several other methods have been proposed and compared \cite{tumminello2007shrinkage,pantaleo2011improved}. From these papers it stems that hierarchical clustering yields better results \cite{tola2008cluster} then other estimators such as shrinkage or RMT-based estimators for correlation matrices of financial time series. The hierarchical clustering filtering procedure first described in \cite{mantegna1999introduction} is illustrated in Figures~\ref{fig:emp_cor} and~\ref{fig:emp_seriated}.
In Figure~\ref{fig:emp_cor}, we display the empirical correlation matrix as estimated on our CDS dataset of $N = 560$ time series of length $T \approx 2500$.
Then, we run a hierarchical clustering algorithm (such as average linkage for example) which gives a re-ordering of the time series, and thus a seriation of the correlation matrix. The re-ordered correlation matrix is displayed in Figure~\ref{fig:emp_seriated} (left). We can now notice its noisy hierarchical correlation structure. According to the hierarchical clustering computed, we can finally filter the correlation coefficients to obtain the correlation matrix displayed in Figure~\ref{fig:emp_seriated} (right).


%\begin{figure}
%\begin{center}
%\includegraphics[scale=0.5]{Fig/seriated_filtered_correlation}
%\caption{The filtered correlation matrix resulting from the method described in \cite{mantegna1999introduction}}\label{fig:emp_filtered}
%\end{center}
%\end{figure}

Mantegna in \cite{mantegna1999hierarchical} and many following papers insist on the hierarchical correlation pattern present in financial time series.
This intrinsic structure may be an explanation to the efficiency of the hierarchical clustering filtering procedure. 
Taking into account other known empirical properties of daily asset returns in liquid financial markets which are well documented in \cite{cont2001empirical}, we do not consider vector autoregression (VAR) modelling and the frequency domain approaches:

\begin{quote}
Mandelbrot
expressed this property by stating that ‘arbitrage tends to
whiten the spectrum of price changes’. This property implies
that traditional tools of signal processing which are based on
second-order properties, in the time domain - autocovariance
analysis, ARMA modelling - or in the spectral domain -
Fourier analysis, linear filtering - cannot distinguish between
asset returns and white noise. This points out the need for
nonlinear measures of dependence in order to characterize the
dependence properties of asset returns. \textit{Excerpt from \cite{cont2001empirical}}
\end{quote}

Now, assuming that data follow this underlying hierarchical correlation model, we may wonder if these clustering procedures are consistent.
Do they always recover the underlying model provided that the time series are long enough? If yes, another interesting point for the practitioner is
knowing the convergence rate. How much data is enough for the result to be reliable? Indeed, since these time series may not be stationary, the practitioner wants to use the shortest time interval possible provided that the results are still relevant.
In the following section, we justify the validity of the clustering approach for the analysis of correlation between financial time series by proving that clustering is statistically consistent in the hierarchical correlation block model. 
We also provide some guidelines to select a good combination of the clustering algorithm, the correlation coefficient, and the minimum number of observations required to obtain meaningful clusters.

\section{On the consistency of clustering correlated random variables}

We show that clustering correlated
random variables from their observations is statistically consistent.
More precisely, when the underlying clusters of correlated random variables satisfy a strong enough separation condition and when there are enough observations, we prove that  many of the celebrated clustering algorithms recover these cluster structures with high probability.
We corroborate our theoretical results with an empirical study of the convergence rates.

Clustering consistency has been widely studied, starting from Hartigan's proof of Single Linkage \cite{hartigan1981consistency} and Pollard's proof of $k$-means consistency \cite{pollard1981strong} to recent work such as the consistency of spectral clustering \cite{von2008consistency}, or modified $k$-means \cite{terada2013strong}, \cite{terada2014strong}.
However, these papers assume that $N$ data points are independently sampled from an underlying probability distribution in dimension $T$ fixed. They show that in the large sample limit, $N \rightarrow \infty$, the clustering structures constructed by the given algorithm converge to a clustering of the whole underlying space.
%This statistical regime corresponds to the \textit{Classical Asymptotics} ($N \rightarrow \infty$, $T$ fixed). 
Much less work has been done to prove consistency of clustering in the \textit{Time Series Asymptotics}, i.e. ($N \rightarrow \infty, T \rightarrow \infty, T/N \rightarrow \infty$) and ($N$ fixed, $T \rightarrow \infty$). We should mention \cite{borysov2014asymptotics} which shows the asymptotic behavior of three hierarchical clustering algorithms, namely Single, Average and Ward Linkage, and their consistency on the task of clustering $N = n + m$ observations from a mixture of two $T$ dimensional Gaussian distributions $\mathcal{N}(\mu_1, \sigma_1^2 I_T)$ and $\mathcal{N}(\mu_2,\sigma_2^2 I_T)$ and 
\cite{ryabko2010clustering}, \cite{khaleghi2012online}, \cite{khaleghi2016consistent} who prove the consistency of $k$-means for clustering processes according only to their distribution. In this work, we show the consistency of clustering $N$ random variables from their $T$ observations according to their observed correlations. The consistency results presented hold for several well-known clustering algorithms, and unlike \cite{borysov2014asymptotics}, we do not assume Gaussian distribution for the random variables, but data assumptions are adjusted to the natural scope of the correlation coefficients (e.g. Gaussian for Pearson correlation, elliptical copula for Kendall tau rank correlation).

\vspace{0.5cm}

\textbf{Notations}
\begin{itemize}
%\item $\rho$, $\rho_S$, $\tau$ the Pearson, Spearman, Kendall correlation respectively.
\item $X_1,\ldots,X_N$ univariate random variables
\item $X_i^t$ is the $t^{\mathrm{th}}$ observation of variable $X_i$
\item $X_i^{(t)}$ is the $t^{\mathrm{th}}$ sorted observation of $X_i$
\item $F_X$ is the cumulative distribution function of $X$
\item $\rho_{ij} = \rho(X_i,X_j)$ correlation between $X_i, X_j$
\item $d_{ij} = d(X_i,X_j)$ distance between $X_i, X_j$
\item $D_{ij} = D(C_i,C_j)$ distance between clusters $C_i, C_j$
\item $P_k = \{C^{(k)}_1,\ldots,C^{(k)}_{l_k}\}$ is a partition of $X_1,\ldots,X_N$
\item $C^{(k)}(X_i)$ denotes the cluster of $X_i$ in partition $P_k$
\item $\Vert \Sigma \Vert_\infty = \max_{ij} \Sigma_{ij}$
\item $X = O_p(k)$ means $X/k$ is stochastically bounded, i.e. $\forall \varepsilon > 0, \exists M > 0, P(|X/k| > M) < \varepsilon$.
\end{itemize}

\subsection{Correlations}

The most common correlation coefficient is the Pearson correlation coefficient defined by
\begin{equation}
\rho(X,Y) = \frac{\mathbb{E}[XY] - \mathbb{E}[X] \mathbb{E}[Y]}{\sqrt{\mathbb{E}[X^2] - \mathbb{E}[X]^2}\sqrt{\mathbb{E}[Y^2] - \mathbb{E}[Y]^2}}
\end{equation}
which can be estimated by
\begin{equation}
\hat{\rho}(X,Y) = \frac{\sum_{t=1}^T (X^t - \overline{X})(Y^t - \overline{Y})}{\sqrt{\sum_{t=1}^T \left(X^t - \overline{X}\right)^2}\sqrt{\sum_{t=1}^T \left(Y^t - \overline{Y}\right)^2}}
\end{equation}
where $\overline{X} = \frac{1}{T}\sum_{t=1}^T X^t$ is the empirical mean of $X$.
This coefficient suffers from several drawbacks: it only measures linear relationship between two variables; it is not robust to noise and may be undefined if the distribution of one of these variables have infinite second moment. More robust correlation coefficients are copula-based dependence measures such as Kendall's tau 
\begin{eqnarray}
\tau(X,Y) & = &  4\int_{0}^{1} \int_{0}^{1} C(u,v) dC(u,v) - 1\\
& = & \mathbb{E}\left[\mathrm{sign}\left((X-\tilde{X})(Y-\tilde{Y})\right) \right]
\end{eqnarray}
where $\tilde{X}$ is an independent copy of $X$,
$C$ is a copula,
and its statistical estimate
\begin{eqnarray}
 \hat{\tau}(X,Y) = \frac{\sum_{1 \leq i < j \leq T} \mathrm{sign}\left( \left(X^{i} - X^{j} \right) \left(Y^{i} - Y^{j} \right) \right)}{{T\choose 2}}
\end{eqnarray}

and Spearman's rho
\begin{eqnarray}
\rho_S(X,Y) & = & 12 \int_{0}^{1} \int_{0}^{1} C(u,v) du dv - 3 \\
& = & 12~\mathbb{E}\left[F_X(X), F_Y(Y) \right] - 3 \\
& = & \rho\left(F_X(X),F_Y(Y)\right)
\end{eqnarray}
and its statistical estimate
\begin{equation}
\hat{\rho}_S(X,Y) = 1 - \frac{6}{T(T^2 - 1)}\sum_{t=1}^T \left(X^{(t)} - Y^{(t)} \right)^2.
\end{equation}
These correlation coefficients are robust to noise (since rank statistics normalize outliers) and invariant to monotonous transformations of the random variables (since copula-based measures benefit from the probability integral transform $F_X(X) \sim \mathcal{U}[0,1]$). 
%If the underlying dependence structure can be represented by an elliptical copula $C$, then the following relations between these correlation coefficients can be derived:
%\begin{eqnarray}
%\rho(X,Y) & = & \sin\left(\frac{\pi}{2}\tau(X,Y)\right) \\
%& = & 2\sin\left(\frac{\pi}{6}\rho_S(X,Y)\right)
%\end{eqnarray}

\subsection{Clustering of correlations: the hierarchical correlation block model}

\begin{figure}[!ht]
\begin{center}
\includegraphics[scale=0.455]{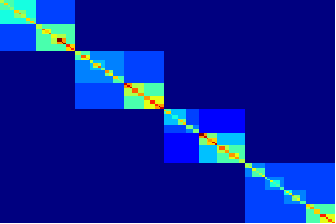}
\includegraphics[scale=0.455]{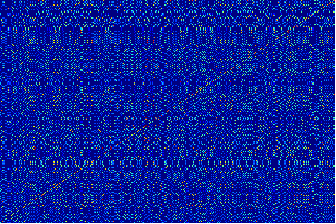}
\caption{(left) hierarchical correlation block model; (right) observed correlation matrix (following the HCBM) identical to the left one up to a permutation of the data}\label{fig:hbm}
\end{center}
\end{figure}

We assume that the $N$ univariate random variables $X_1,\ldots,X_N$ follow a Hierarchical Correlation Block Model (HCBM). This model consists in correlation matrices having a hierarchical block structure \cite{balakrishnan2011noise}, \cite{krishnamurthy2012efficient}. Each block corresponds to a correlation cluster that we want to recover with a clustering algorithm. In Fig.~\ref{fig:hbm}, we display a correlation matrix from the HCBM. Notice that in practice one does not observe the hierarchical block diagonal structure displayed in the left picture, but a correlation matrix similar to the one displayed in the right picture which is identical to the left one up to a permutation of the data.
The HCBM defines a set of nested partitions $\mathcal{P} = \{P_0 \supseteq P_1 \supseteq \ldots \supseteq P_h\}$ for some $h \in [1,N]$, where $P_0$ is the trivial partition, the partitions $P_k = \{C^{(k)}_1,\ldots,C^{(k)}_{l_k}\}$, and $\sqcup_{i = 1}^{l_k} C^{(k)}_i = \{X_1,\ldots,X_N\}$. For all $1 \leq k \leq h$, we define $\underline{\rho}_k$ and  $\overline{\rho}_k$ such that for all $1 \leq i,j \leq N$, we have $\underline{\rho}_k \leq \rho_{ij} \leq \overline{\rho}_k$ when $C^{(k)}(X_i) = C^{(k)}(X_j)$ and $C^{(k+1)}(X_i) \neq C^{(k+1)}(X_j)$, i.e. $\underline{\rho}_k$ and $\overline{\rho}_k$ are the minimum and maximum correlation respectively within all the clusters $ C^{(k)}_i$ in the partition $P_k$ at depth $k$. In order to have a proper nested correlation hierarchy, we must have $\overline{\rho}_k < \underline{\rho}_{k+1}$ for all $k$.

Without loss of generality and for ease of demonstration we will consider the one-level HCBM with $K$ blocks of size $n_1,\ldots,n_K$ such that $\sum_{i=1}^K n_i = N$. We explain later how to extend the results to the general HCBM. Since clustering methods usually require a distance
matrix as input, we also consider the corresponding distance matrix with coefficients $d_{ij} = \frac{1-\rho_{ij}}{2}$, where $0 < \rho_{ij} < 1$ is a correlation coefficient (Pearson, Spearman, Kendall).

\subsection{Clustering methods}

Many paradigms exist in the literature for clustering data. We consider in this work only hard (in opposition to soft) clustering methods, i.e. algorithms producing partitions of the data (in opposition to methods assigning several clusters to a given data point). 
Within the hard clustering family, we can classify for instance these algorithms in hierarchical clustering methods (yielding nested partitions of the data) and flat clustering methods (yielding a single partition) such as $k$-means.

We will consider the infinite Lance-Williams family which further subdivides the hierarchical clustering since many of the popular algorithms such as Single Linkage, Complete Linkage, Average Linkage (UPGMA), McQuitty's Linkage (WPGMA), Median Linkage (WPGMC), Centroid Linkage (UPGMC), and Ward's method are members of this family (cf. Table~\ref{LWparams}). It will allow us a more concise and unified treatment of the consistency proofs for these algorithms. Interesting and recently designed hierarchical agglomerative clustering algorithms such as Hausdorff Linkage \cite{basalto2007hausdorff} and Minimax Linkage \cite{ao2005clustag} do not belong to this family \cite{bien2011hierarchical}, but their linkage functions share a convenient property for cluster separability. 

\begin{table}[]
\centering
\caption{Many well-known hierarchical agglomerative clustering algorithms are members of the Lance-Williams family, i.e. the distance between clusters can be written: \newline $D(C_i \cup C_j, C_k) = \alpha_i D_{ik} + \alpha_j D_{jk} + \beta D_{ij} + \gamma |D_{ik} - D_{jk}|$ \cite{murtagh2011methods}}
\label{LWparams}
\begin{tabular}{l|c|c|c|}
\cline{2-4}
  & $\alpha_i$ & $\beta$ & $\gamma$   \\
  \hline
\multicolumn{1}{|l|}{Single} &  1/2  &  0      &     -1/2     \\
\hline
\multicolumn{1}{|l|}{Complete} &   1/2    &  0      &  1/2        \\
\hline
\multicolumn{1}{|l|}{Average}  &  $\frac{|C_i|}{|C_i| + |C_j|}$      &   0    &  0  \\
  \hline
\multicolumn{1}{|l|}{McQuitty}  &   1/2     &  0     & 0    \\
  \hline
\multicolumn{1}{|l|}{Median}  &   1/2     & -1/4      &  0  \\
  \hline
\multicolumn{1}{|l|}{Centroid}  &   $\frac{|C_i|}{|C_i| + |C_j|}$     & $- \frac{|C_i| |C_j|}{(|C_i|+ |C_j|)^2}$       &   0 \\
  \hline
\multicolumn{1}{|l|}{Ward}  &  $\frac{|C_i|+|C_k|}{|C_i|+|C_j|+|C_k|}$  & $- \frac{|C_k|}{C_i|+|C_j|+|C_k|}$       &  0  \\
  \hline
\end{tabular}
\end{table}

\subsection{Separability conditions for clustering}

%It is easy to verify that the theoretical distance matrix of the HCBM model verifies the separability condition between clusters for all the algorithms we study whenever the intra cluster distance are smaller than the inter cluster distance. However what we can observe is a noisy empirical version of $d$: $\hat{d}$. In this section we derive sufficient separability condition between clusters for the clustering algorithm to find the true partition.
In our context the distances between the points we want to cluster are random and defined by the estimated correlations. However by definition of the HCBM, each point $X_i$ belongs to exactly one cluster $C^{(k)}(X_i)$ at a given depth $k$, and we want to know under which condition on the distance matrix we will find the correct clusters defined by $P_k$. We call these conditions the separability conditions.
A separability condition for the points $X_1,\ldots,X_N$ is a condition on the distance matrix of these points such that if we apply a clustering procedure whose input is the distance matrix, then the algorithm yields the correct clustering $P_k = \{C^{(k)}_1,\ldots,C^{(k)}_{l_k} \}$, for all $k$.
For example, for $\{X_1,X_2,X_3\}$ if we have $C(X_1) = C(X_2) \neq C(X_3)$ in the one-level two-block HCBM, then a separability condition is $d_{1,2} < d_{1,3}$ and $d_{1,2} < d_{2,3}$.
%In the nested partition model, it is a condition such that the algorithm finds all the nested partitions.

Separability conditions are deterministic and depend on the algorithm used for clustering. They are generic in the sense that for any sets of points that satisfy the condition the algorithm will separate them in the correct clusters. In the Lance-Williams algorithm framework \cite{chen1996space}, they are closely related to ``space conserving'' properties of the algorithm and in particular on the way the distances between clusters change during the clustering process.

In \cite{chen1996space}, the authors define what they call a semi-space-conserving algorithm.

\textbf{Semi-space-conserving algorithms \cite{chen1996space}}
 
An algorithm is semi-space-conserving if for all clusters $C_i$, $C_j$, and $C_k$,
$$
 D(C_i\cup C_j,C_k) \in  \left[\min(D_{ik},D_{jk}), \max(D_{ik},D_{jk}) \right]
$$

Among the Lance-Williams algorithms we study here, Single, Complete, Average and McQuitty algorithms are semi-space-conserving. Although Chen and Van Ness only considered Lance-Williams algorithms the definition of a space conserving algorithm is useful for any agglomerative hierarchical algorithm. An alternative formulation of the semi-space-conserving property is:

\textbf{Space-conserving algorithms.}
A linkage agglomerative hierarchical algorithm is space-conserving if $\displaystyle D_{ij}\in\left[\min_{x\in C_i,y\in C_j} d(x,y),\max_{x\in C_i,y\in C_j} d(x,y)\right]$.

Such an algorithm does not ``distort" the space when points are clustered which makes the sufficient separability condition easier to get. For these algorithms the separability condition does not depend on the size of the clusters.

\vspace{0.2cm}
The following two propositions are easy to verify.

\textit{Proposition.} The semi-space-conserving Lance-Williams algorithms are space-conserving.

\textit{Proposition.} Minimax linkage and Hausdorff linkage are space-conserving.

\vspace{0.2cm}

For space-conserving algorithms we can now state a sufficient separability condition on the  distance matrix.

\textit{Proposition.}
The following condition is a separability condition for space-conserving algorithms:
\begin{equation}
\tag{S1}
\max_{\substack{1 \leq i,j \leq N\\C(i) = C(j)}} d(X_i,X_j) ~< \min_{\substack{1 \leq i,j \leq N\\C(i) \neq C(j)}} d(X_i,X_j)
\label{sep_clust}
\end{equation}
The maximum distance is taken over any two points in a same cluster (intra) and the minimum over any two points in different clusters (inter).

\proof Consider the set $\{d_{ij}^s\}$ of distances between clusters after $s$ steps of the clustering algorithm (therefore $\{d_{ij}^0\}$ is the initial set of distances between the points). Denote $\{d_{inter}^s\}$ (resp. $\{d_{intra}^s\}$) the sets of distances between subclusters belonging to different clusters (resp. the same cluster) at step $s$.
If the separability condition is satisfied then we have the following inequalities:

\begin{equation}
  \tag{S2}
\min d^0_{intra}\leq\max d^0_{intra}<\min d^0_{inter}\leq\max d^0_{inter}
%\underline{d}_1\leq d_{i,j}^k\leq \overline{d}_1<\underline{d}_0\leq d_{i,j}^k\leq \overline{d}_0
\label{sep}
\end{equation}

Then the separability condition implies that the separability condition \ref{sep} is verified for all step $s$ because after each step the updated intra distances are in the convex hull of the intra distances of the previous step and the same is true for the inter distances. Moreover since \ref{sep} is verified after each step, the algorithm never links points from different clusters and the proposition entails.~$\qed$

\subsection{Concentration bounds for the correlation matrix}

We have determined configurations of points such that the clustering algorithm will find the right partition. The proof of the consistency now relies on showing that these configurations are likely. In fact the probability that our points fall in these configurations goes to $1$ as $T\to\infty$. 

\vspace{0.2cm}
The precise definition of what we mean by consistency of an algorithm is the following:

\textbf{Consistency of a clustering algorithm.}
Let $(X_1^t,\ldots,X_N^t)$, $t=1,\ldots,T$, be $N$ univariate random variables observed $T$ times. A clustering algorithm $\mathcal{A}$ is consistent with respect to the Hierarchical Correlation Block Model (HCBM) defining a set of nested partitions $\mathcal{P}$ if the probability that the algorithm $\mathcal{A}$ recovers all the partitions in $\mathcal{P}$ converges to $1$ when $T \rightarrow \infty$.

\vspace{0.2cm}
We now get explicit lower bounds on the probability of finding the right clusters with the clustering algorithms using concentration bounds on the empirical correlation matrix.
 
As we have seen in the previous section the correct clustering can be ensured if the estimated correlation matrix verifies some separability condition. This condition can be guaranteed by requiring the error on each entry of the matrix $\hat{R}_T$ to be smaller than the contrast, i.e. $\frac{\underline{\rho}_1-\overline{\rho}_0}{2}$, on the theoretical matrix $R$. In general the error on the matrix $\hat{R}_T$ is of the order $\Vert R-\hat{R}_T\Vert_{\infty}=O_{P}\left(\sqrt{\frac{\log N}{T}}\right)$ and thus, if $T \gg \log(N)$ then the clustering will find the correct partition.

%\textbf{Concentration bound on the Pearson correlation matrix $\Sigma$}
%
%Let $X^t$, $t=1,\ldots,T$, be $T$ independent realizations of the $N$-dimensional Gaussian distribution with covariance matrix $\Sigma$. Let $\hat{\Sigma}_T$ be the empirical Pearson correlation matrix. We obtain
%\begin{equation}
%\mathbb{P}\left(\Vert\hat{\Sigma}_T-\Sigma\Vert_{\infty}\leq \epsilon\right) \geq 1-2N^2e^{-I(\epsilon)T}
%\end{equation}
%where $I(\epsilon) = \begin{cases}
%\ln\left(\frac{1-\epsilon\underline{\rho}_0}{\sqrt{(1-\underline{\rho}_0^2)(1-\epsilon^2)}}\right), & 0<\epsilon<1\\
%+\infty, & \epsilon\geq 1 
%\end{cases}
%$.

%\textbf{Concentration bound on the Spearman's rho correlation matrix $R$}
%
%Let $X^t$, $t=1,\ldots,T$, be $T$ independent realizations of the $N$-dimensional  monotone-nonparanormal distribution \cite{liu2009nonparanormal}, i.e. a $N$-dimensional Gaussian distribution with covariance $\Sigma$ after some monotone transformations of the variables. Then, we have 
%\begin{equation}
%\mathbb{P}\left(\Vert\hat{R}_T-R\Vert_{\infty}\leq \epsilon\right) \geq 1-2N^2e^{-\frac{T}{18}\epsilon^2}
%\end{equation}
%whenever $T\geq\frac{10}{\epsilon}$.

Results and proofs are the object of an upcoming publication. Below we only give the results for the Kendall's tau coefficient, but Spearman's bound is similar.

\textbf{Concentration bound on the Kendall's tau correlation matrix $U$}

Let $X^t$, $t=1,\ldots,T$, be $T$ independent realizations of a $N$-dimensional distribution having elliptical copula and any margins. We have
\begin{equation}
\mathbb{P}\left(\Vert\hat{U}_T-U\Vert_{\infty}\leq \epsilon\right) \geq 1-2N^2e^{-\frac{T}{8}\epsilon^2}.
\end{equation}

The lower bound on the probability of success now follows by requiring that the error on the estimated correlation matrix is small enough. Moreover $\rho$ is taken to be a generic correlation and $\Sigma$ the corresponding generic correlation matrix. %Getting the specific bound for each correlation coefficient is just a question of replacing a generic constant $K$ by the specific one for each coefficient.

\textbf{Space-conserving algorithms}

The separability condition is satisfied if $\Vert \Sigma-\hat{\Sigma}\Vert_\infty<\frac{\underline{\rho}_1-\overline{\rho}_0}{2}$. Therefore with probability at least
\begin{equation}
1-2N^2e^{-\frac{T\left(\underline{\rho}_1-\overline{\rho}_0\right)^2}{32}}
\end{equation}
for Kendall correlation,
the algorithm finds the correct partition.

Therefore we obtain consistency for the presented algorithms with respect to the one-level HCBM.

\subsection{From the one-level to the general HCBM}

To go from the one-level HCBM to the general case we need to get a separability condition on the nested partition model. For space-conserving algorithms, this is done by requiring the corresponding separability condition for each level of the hierarchy. 

For all $1 \leq k \leq h$, we define $\underline{d}_k$ and $\overline{d}_k$ such that for all $1 \leq i,j \leq N$, we have $\underline{d}_k \leq d_{ij} \leq \overline{d}_k$ when $C^{(k)}(X_i) = C^{(k)}(X_j)$ and $C^{(k+1)}(X_i) \neq C^{(k+1)}(X_j)$. Notice that $\underline{d}_k = (1-\overline{\rho}_{k})/2$ and $\overline{d}_k = (1-\underline{\rho}_k)/2$.

\textbf{Separability condition for space-conserving algorithms in the case of nested partitions.} The separability condition reads:
$$\overline{d}_h < \underline{d}_{h-1} < \ldots < \overline{d}_{k+1} < \underline{d}_{k} < \ldots < \underline{d}_{1}.$$

This condition can be guaranteed by requiring the error on each entry of the matrix $\hat{\Sigma}$ to be smaller than the lowest contrast.
Therefore the maximum error we can have for space-conserving algorithms on the correlation matrix is $$\Vert \Sigma-\hat{\Sigma}\Vert_\infty<\min_{k}\frac{\underline{\rho}_{k+1}-\overline{\rho}_k}{2}.$$

We finally obtain consistency for the presented algorithms with respect to the HCBM from the previous concentration bounds.

\subsection{Empirical rates of convergence}

Researchers have used from 30 days to several years of daily returns as source data for clustering financial time series based on their correlations.
How long should the time series be? If too short, the clusters found can be spurious; if too long, dynamics can be smoothed out. A practical methodology to address this issue can be found in \cite{marti2016clustering}.

For illustration purpose, we consider the simple case where we have two correlation blocks $C_1$ and $C_2$. The correlation within the block $C_1$ is $\rho$ and within the block $C_2$ is $2\rho$ and both blocks are independent. $C_2$ counts for $70\%$ of the $N$ points.
 The underlying correlation matrix is thus of the form:

$$
\footnotesize
\arraycolsep=3pt
\medmuskip = 1mu % default: 4mu plus 2mu minus 4mu
     \begin{pmatrix}
		\textcolor{darkgreen}{1}        &  \textcolor{darkgreen}{2\rho}    &  \textcolor{darkgreen}{\cdots}  &  \textcolor{darkgreen}{2\rho}    & \textcolor{red}{0}        &  \textcolor{red}{\cdots}  &  \textcolor{red}{\cdots}  & \textcolor{red}{\cdots}        & \textcolor{red}{\cdots} &\textcolor{red}{0} \\
		 \textcolor{darkgreen}{2\rho}    &  \textcolor{darkgreen}{\ddots}  &  \textcolor{darkgreen}{\ddots}  &  \textcolor{darkgreen}{\vdots}  &  \textcolor{red}{\vdots}  & \textcolor{red}{\ddots}         &     \textcolor{red}{\ddots}       &   \textcolor{red}{\ddots}   &\textcolor{red}{\ddots}   &\textcolor{red}{\vdots} \\
		 \textcolor{darkgreen}{\vdots}  &  \textcolor{darkgreen}{\ddots}  &  \textcolor{darkgreen}{\ddots}  &  \textcolor{darkgreen}{2\rho }   &  \textcolor{red}{\vdots}  &     \textcolor{red}{\ddots}       &     \textcolor{red}{\ddots}       &  \textcolor{red}{\ddots}    &\textcolor{red}{\ddots}   & \textcolor{red}{\vdots}\\
		 \textcolor{darkgreen}{2\rho}    &  \textcolor{darkgreen}{\cdots}  &  \textcolor{darkgreen}{2\rho}    & \textcolor{darkgreen}{1}		  & \textcolor{red}{0}        &  \textcolor{red}{\cdots}  &  \textcolor{red}{\cdots}  & \textcolor{red}{\cdots}        & \textcolor{red}{\cdots}  &\textcolor{red}{0} \\
		\textcolor{red}{0}        &  \textcolor{red}{\cdots}  &  \textcolor{red}{\cdots}  & \textcolor{red}{0}        & \textcolor{darkblue}{1}        &  \textcolor{darkblue}{\rho}    &  \textcolor{darkblue}{\cdots}  &  \textcolor{darkblue}{\cdots}    & \textcolor{darkblue}{\cdots} & \textcolor{darkblue}{\rho} \\
		 \textcolor{red}{\vdots}  &    \textcolor{red}{\ddots}        &    \textcolor{red}{\ddots}        &  \textcolor{red}{\vdots}  &  \textcolor{darkblue}{\rho}    &  \textcolor{darkblue}{\ddots}  &  \textcolor{darkblue}{\ddots}  &  \textcolor{darkblue}{\ddots}  &\textcolor{darkblue}{\ddots} & \textcolor{darkblue}{\vdots}\\
		 \textcolor{red}{\vdots}  &     \textcolor{red}{\ddots}       &      \textcolor{red}{\ddots}      &  \textcolor{red}{\vdots}  &  \textcolor{darkblue}{\vdots}  &  \textcolor{darkblue}{\ddots}  &  \textcolor{darkblue}{\ddots}  &  \textcolor{darkblue}{\ddots}    &\textcolor{darkblue}{\ddots} & \textcolor{darkblue}{\vdots}\\
		\textcolor{red}{\vdots}        &  \textcolor{red}{\ddots}    &  \textcolor{red}{\ddots}    & \textcolor{red}{\vdots}        & \textcolor{darkblue}{\vdots}   &  \textcolor{darkblue}{\ddots} &    \textcolor{darkblue}{\ddots}  & \textcolor{darkblue}{\ddots}&\textcolor{darkblue}{\ddots} &\textcolor{darkblue}{\vdots} \\
			\textcolor{red}{\vdots}        & \textcolor{red}{\ddots}     & \textcolor{red}{\ddots}     & \textcolor{red}{\vdots}        &  \textcolor{darkblue}{\vdots}    & \textcolor{darkblue}{\ddots}   &   \textcolor{darkblue}{\ddots}   & \textcolor{darkblue}{\ddots} &\textcolor{darkblue}{\ddots} & \textcolor{darkblue}{\rho}\\
				\textcolor{red}{0}        &  \textcolor{red}{\cdots}  &  \textcolor{red}{\cdots}  & \textcolor{red}{0}        &  \textcolor{darkblue}{\rho}    &  \textcolor{darkblue}{\cdots}  &  \textcolor{darkblue}{\cdots}    & \textcolor{darkblue}{\cdots}& \textcolor{darkblue}{\rho}&\textcolor{darkblue}{1} \\
		
\end{pmatrix}$$

We then simulate Gaussian and Student (with $\nu = 3$ degrees of freedom, i.e. heavy-tailed) random vectors, create the different correlation matrices and cluster with these matrices using the Ward, Single, Complete and Average Linkage algorithms. We then count the number of success of these clustering procedures, i.e. finding the correct partition, over $100$ trials. This experiment has been done for the two sets of parameters $(N,T)$ and $(\rho,T)$. We produce the heat maps (relative to the number of successes) for these different experiments.

\subsubsection{$(N,T)$ experiments.}

In this first experiment $\rho$ is fixed at $0.1$ and we do the clustering procedure for different values of $N$ and $T$.

\begin{figure}[!ht]
\begin{center}
\includegraphics[width=0.49\columnwidth]{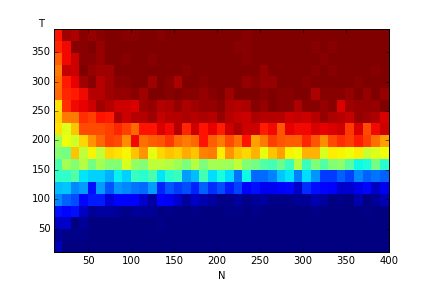}
\includegraphics[width=0.49\columnwidth]{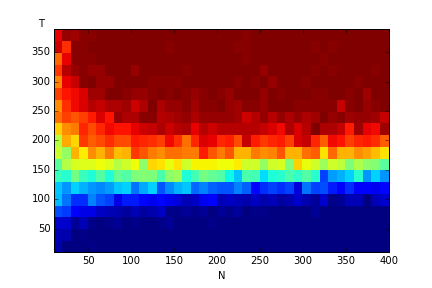}
\caption{Single Linkage applied on (left) Spearman dissimilarity, (right) Pearson dissimilarity; the $x$ axis is $N=10\ldots400$, the $y$ axis is $T=10\ldots390$. }\label{heatmap1}
\end{center}
\end{figure}

%\begin{figure}[!ht]
%\begin{center}
%\includegraphics[width=0.49\columnwidth]{Fig/heatmap_spearman_ward_NT}
%\includegraphics[width=0.49\columnwidth]{Fig/heatmap_pearson_ward_NT}
%\caption{Ward Linkage applied on (left) Spearman dissimilarity, (right) Pearson dissimilarity; the $x$ axis is $N=10\ldots400$, the $y$ axis is $T=10\ldots390$.}\label{heatmap2}
%\end{center}
%\end{figure}

%\begin{figure}[!ht]
%\begin{center}
%\includegraphics[scale=0.45]{heatmap_spearman_complete_NT}
%\includegraphics[scale=0.45]{heatmap_pearson_complete_NT}
%\caption{Complete Linkage applied on (left) Spearman dissimilarity, (right) Pearson %dissimilarity; the $x$ axis is $N=10\ldots400$, the $y$ axis is %$T=10\ldots390$.}\label{heatmap3}
%\end{center}
%\end{figure}

%\begin{figure}[!ht]
%\begin{center}
%\includegraphics[scale=0.45]{heatmap_spearman_average_NT}
%\includegraphics[scale=0.45]{heatmap_pearson_average_NT}
%\caption{Average Linkage applied on (left) Spearman dissimilarity, (right) Pearson %dissimilarity; the $x$ axis is $N=10\ldots400$, the $y$ axis is %$T=10\ldots390$.}\label{heatmap4}
%\end{center}
%\end{figure}

As one can see in Fig.~\ref{heatmap1}, there is a ``transition" area between zones with probability almost $1$ and almost $0$ of finding the right clusters. The absolute level of this transition zone depends on the clustering algorithm. What we can see in these examples is that the dependence in $T$ is much quicker than in $N$ and that in fact in our sample for $N>100$ there is little dependence in $N$.

For moderately sized group of points, typically $100 \leq N \leq 400$, we can deduce that for $T \geq 250$ all of the clustering algorithms find the correct partition in the HCBM model with very high probability (cf. Table~\ref{tab_success}).
\begin{table}[]
\centering
\caption{Number of success out of 100 trials for $T=250$ and $N=400$}\label{tab_success}
\begin{tabular}{|l|c|c|c|}
  \hline
    & single & average & complete \\
  \hline
  Pearson  & 98 & 98 & 99 \\
  \hline
  Spearman & 95 & 99 & 100 \\
  \hline
\end{tabular}
\end{table}

\subsubsection{$(\rho,T)$ experiments.}

For the $(\rho,T)$ experiments, we made two different sets of experiments both with the Spearman correlation matrix and the Pearson correlation matrix. One with Gaussian random variables and the other with multivariate Student variables (with $\nu = 3$ degrees of freedom) which exhibit fatter tails.

\begin{figure}[!ht]
\begin{center}
\includegraphics[width=0.49\columnwidth]{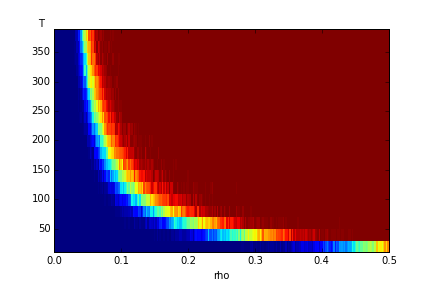}
\includegraphics[width=0.49\columnwidth]{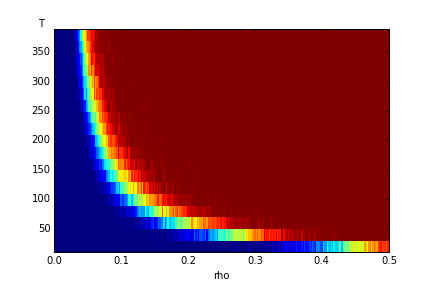}
\caption{Gaussian case for Spearman (left) and Pearson (right) and for the Average Linkage. The $x$ axis is $\rho=0\ldots0.5$ and the $y$ axis is $T=10\ldots390$.}\label{heatmap5}
\end{center}
\end{figure}

\begin{figure}[!ht]
\begin{center}
\includegraphics[width=0.49\columnwidth]{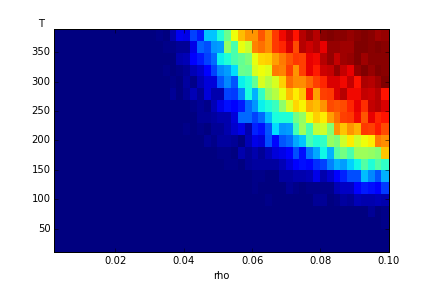}
\includegraphics[width=0.49\columnwidth]{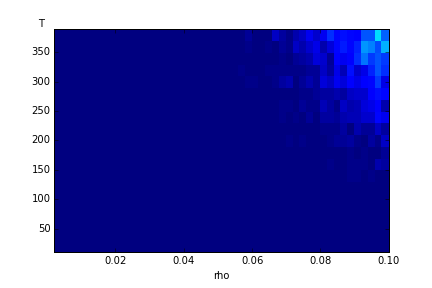}
\caption{Student case for Spearman (left) and Pearson (right) and for the Average Linkage. The $x$ axis is $\rho=0\ldots0.1$ and the $y$ axis is $T=10\ldots390$.}\label{fat_tails}
\end{center}
\end{figure}

As expected with the Student distribution, the Pearson correlation coefficient is not robust to fatter tails and the clustering rate of success is much lower than in the Gaussian case (Fig.~\ref{heatmap5}) as it can be seen in Fig.~\ref{fat_tails}. 
%What is very interesting is that the curve delimiting the area with almost perfect success rate is consistent with a decreasing exponential shape suggesting that we are close to capturing the real $\rho$ asymptotic theoretically for the space-conserving algorithms.

Concretely, our results suggest that for properly clustering $N \simeq 400$ correlated financial time series, the practitioner should need $T \geq 250$, i.e. at least a year of daily prices. We also advise to measure correlation with the Kendall coefficient since 
\begin{itemize}
\item more generic: Kendall can be used with any elliptical copula and any margins,
\item unbiased (unlike Spearman),
\item faster convergence rate (than Spearman corrected from the bias),
\item can be computed efficiently in $O(T\log T)$ vs $O(T\log T)$ for Spearman and $O(T)$ for Pearson.
\end{itemize}

%(i) robust to outliers alike Kendall's one, but unlike the Pearson coefficient which is sensitive to heavy-tailed data, (ii) faster computational complexity than the Kendall's one. 
%Any algorithms of the above can be used, but prefer Average or Complete Linkage for a faster convergence rate in practice, and a resilience to a noisy %underlying HCBM structure.
%One should retrieve meaningful clusters provided that $\rho \geq 0.2$ which is realistic on financial markets.

 We notice that there are isoquants of clustering accuracy for many sets of parameters, e.g. $(N,T)$, $(\rho,T)$. Such isoquants are displayed in Figure~\ref{heatmap5}. Further work may aim at characterizing these curves. We can also observe in Figure~\ref{heatmap5} that for $\rho \leq 0.08$, the critical value for $T$ explodes. It would be interesting to determine this asymptotics as $\rho$ tends to 0.

However it is observed that clusters are unstable (with respect to the clustering method \cite{lemieux2014clustering}, and with respect to the clustering distance \cite{marti2015proposal}).
It suggests that information present in the financial time series may not be summarized by cross-correlation only, even under the random walk hypothesis \cite{donnat2016toward}.

\section{Beyond correlation: toward a geometry of the (copula, margins) representation}

In this section, we provide avenues for tackling shortcoming (i) when clustering, i.e. the assumption that assets' returns are following a Gaussian multivariate distribution.
If the assets' returns are not jointly Gaussian distributed, then the variance-covariance matrix does not capture their dependence: linear (Pearson) correlation measures a mixed information of linear dependence and marginals' effect on it. Few `outliers' returns of some assets due to specific events or erroneous values in the data (i.e. tail-realizations from an heavy-tailed distribution) can lower drastically the measured correlation making one to believe that assets are weakly correlated and that investing in them is a diversified investment. Besides, even if several assets are perfectly `correlated', one may still want to discriminate between assets that have high volatility from those of low volatility while doing clustering or risk analysis.

\subsection{A first approach with $N$ univariate time series}

A naive but often used distance between random variables to measure similarity and to perform clustering is the $L_2$ distance $\mathbb{E}[ (X-Y)^2 ]$.
Yet, this distance is not suited to our task. 
\newtheorem{exmp}{Example}
\begin{exmp}[Distance $L_2$ between two Gaussians]
Let $(X,Y)$ be a bivariate Gaussian vector, with $X \sim \mathcal{N}(\mu_X,\sigma_X^2)$, $Y \sim \mathcal{N}(\mu_Y,\sigma_Y^2)$ and whose correlation is $\rho(X,Y) \in [-1,1]$. We obtain $\mathbb{E}[ (X-Y)^2 ] = (\mu_X - \mu_Y)^2 + (\sigma_X - \sigma_Y)^2 + 2\sigma_X \sigma_Y ( 1 - \rho(X,Y) )$. Now, consider the following values for correlation:
\begin{itemize}
\item $\rho(X,Y) = 0$, so $\mathbb{E}[ (X-Y)^2 ] = (\mu_X - \mu_Y)^2 + \sigma_X^2 + \sigma_Y^2$.
The two variables are independent (since uncorrelated and jointly normally distributed), thus we must discriminate on distribution information. Assume $\mu_X = \mu_Y$ and $\sigma_X = \sigma_Y$. For $\sigma_X = \sigma_Y \gg 1$, we obtain $\mathbb{E}[ (X-Y)^2 ] \gg 1$ instead of the distance $0$, expected from comparing two equal Gaussians.
\item $\rho(X,Y) = 1$, so $\mathbb{E}[ (X-Y)^2 ] = (\mu_X - \mu_Y)^2 + (\sigma_X - \sigma_Y)^2$.
Since the variables are perfectly correlated, we must discriminate on distributions. We actually compare them with a $L_2$ metric on the mean $\times$ standard deviation half-plane. However, this is not an appropriate geometry for comparing two Gaussians \cite{costa2014fisher}. For instance, if $\sigma_X = \sigma_Y = \sigma$, we find $\mathbb{E}[ (X-Y)^2 ] = (\mu_X - \mu_Y)^2$ for any values of $\sigma$. As $\sigma$ grows, probability attached by the two Gaussians to a given interval grows similar (cf. Fig.~\ref{fig:equidist_Gaussians}), yet this increasing similarity is not taken into account by this $L_2$ distance.
\begin{figure}
\vskip 0.2in
\begin{center}
\includegraphics[width=0.5\linewidth]{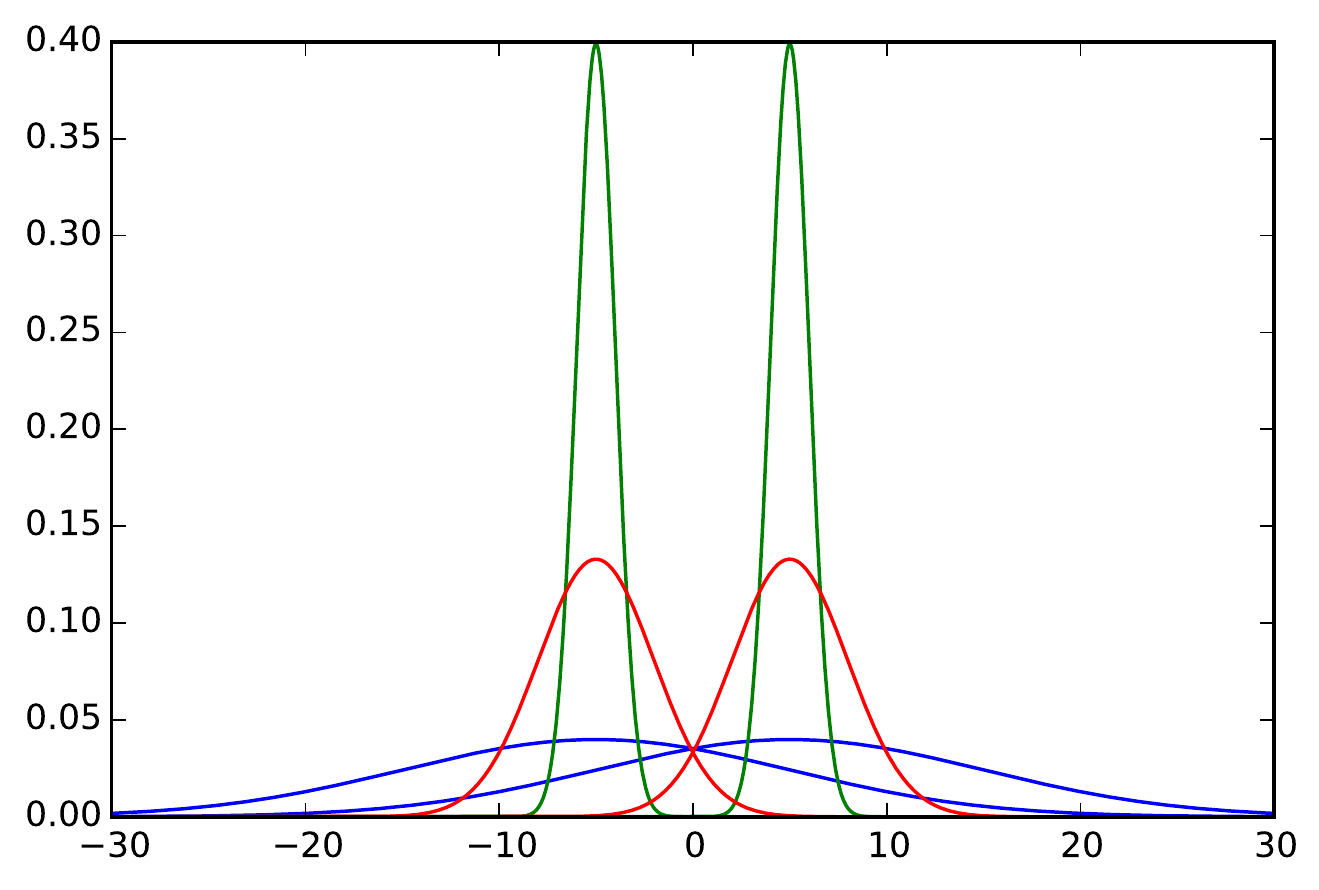}
\caption{Probability density functions of Gaussians $\mathcal{N}(-5,1)$ and $\mathcal{N}(5,1)$ (in green), Gaussians $\mathcal{N}(-5,3)$ and $\mathcal{N}(5,3)$ (in red), and Gaussians $\mathcal{N}(-5,10)$ and $\mathcal{N}(5,10)$ (in blue). Green, red and blue Gaussians are equidistant using $L_2$ geometry on the parameter space $(\mu,\sigma)$.}
\label{fig:equidist_Gaussians}
\end{center}
\vskip -0.2in
\end{figure}
\end{itemize}
\end{exmp}
$\mathbb{E}[(X-Y)^2]$ considers both dependence and distribution information of the random variables, but not in a relevant way with respect to our task.
Our purpose is to introduce a new data representation and a suitable distance which takes into account both distributional proximities and joint behaviours.

Let $(\Omega,\mathcal{F},\mathbb{P})$ be a probability space. $\Omega$ is the sample space, $\mathcal{F}$ is the $\sigma$-algebra of events, and $\mathbb{P}$ is the probability measure.
Let $\mathcal{V}$ be the space of all continuous real-valued random variables defined on $(\Omega,\mathcal{F},\mathbb{P})$.
Let $\mathcal{U}$ be the space of random variables following a uniform distribution on $[0,1]$ and $\mathcal{G}$ be the space of absolutely continuous cumulative distribution functions (cdf).

\vspace{0.2cm}
\textbf{The copula transform}
Let $X = (X_1,\ldots,X_N) \in \mathcal{V}^N$ be a random vector with cdfs $G_X = (G_{X_1},\ldots,G_{X_N}) \in \mathcal{G}^N$. The random vector $G_X(X) = (G_{X_1}(X_1),\ldots,G_{X_N}(X_N)) \in \mathcal{U}^N$ is known as the copula transform.

\vspace{0.2cm}
\textbf{Uniform margins of the copula transform}
$G_{X_i}(X_i)$, $1 \leq i \leq N$, are uniformly distributed on $[0,1]$.

\begin{proof}
$x = G_{X_i}(G_{X_i}^{-1}(x)) = \mathbb{P}(X_i \leq G_{X_i}^{-1}(x)) = \mathbb{P}(G_{X_i}(X_i) \leq x)$.
\end{proof}
We define the following representation of random vectors  that actually splits the joint behaviours of the marginal variables from their distributional information.

\vspace{0.2cm}
\textbf{Dependence $\oplus$ distribution space projection.}
Let $\mathcal{T}$ be a mapping which transforms $X = (X_1,\ldots,X_N)$ into its generic representation, an element of $ \mathcal{U}^N\times\mathcal{G}^N$ representing $X$, defined as follow
\begin{eqnarray}
\mathcal{T}:\mathcal{V}^N & \rightarrow & \mathcal{U}^N\times\mathcal{G}^N \\ 
X & \mapsto & (G_X(X),G_X). \nonumber
\end{eqnarray}

%where $G_X = (G_{X_1},\ldots,G_{X_N})$, and $G_{X_i}$ being the cumulative distribution function of $X_i$.

$\mathcal{T}$ is a bijection.

\begin{figure*}
\vskip 0.2in
\begin{center}
\centerline{\includegraphics[width=0.3\textwidth]{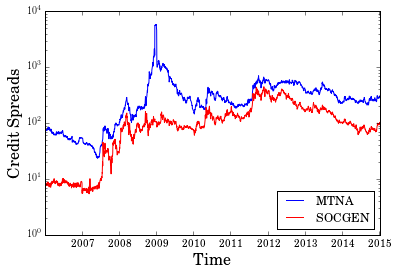}
\includegraphics[width=0.03\textwidth]{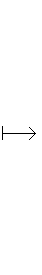}
\includegraphics[width=0.3\textwidth]{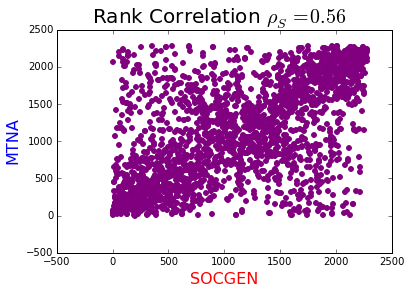} \includegraphics[width=0.03\textwidth]{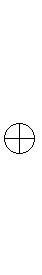} 
\includegraphics[width=0.29\textwidth]{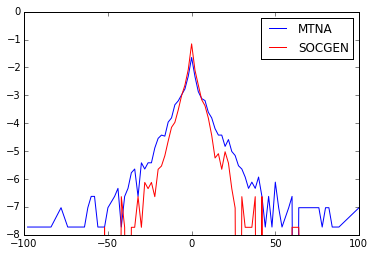}
}
\caption{ArcelorMittal and Soci\'et\'e g\'en\'erale prices ($T$ observations  $(X_1^t,X_2^t)_{t=1}^T$ from $(X_1,X_2) \in \mathcal{V}^2$) are projected on dependence $\oplus$ distribution space; $(G_{X_1}(X_1),G_{X_2}(X_2)) \in \mathcal{U}^2$ encode the dependence between $X_1$ and $X_2$ (a perfect correlation would be represented by a sharp diagonal on the scatterplot); $(G_{X_1},G_{X_2})$ are the margins (their log-densities are displayed above), notice their heavy-tailed exponential distribution (especially for ArcelorMittal).}
\label{GNPR_projection}
\end{center}
\vskip -0.2in
\end{figure*}

\begin{proof}
$\mathcal{T}$ is surjective as any element $(U,G) \in \mathcal{U}^N\times\mathcal{G}^N$ has the fiber $G^{-1}(U)$. $\mathcal{T}$ is injective as $(U_{1},G_{1}) = (U_{2},G_{2})$ \textit{a.s.} in  $\mathcal{U}^N\times\mathcal{G}^N$ implies that they have the same cdf $G=G_{1}=G_{2}$ and since $U_{1}=U_{2}$ \textit{a.s.}, it follows that $G^{-1}(U_{1})=G^{-1}(U_{2})$ \textit{a.s.}
\end{proof}
This result replicates the seminal result of copula theory, namely Sklar's theorem \cite{sklar1959fonctions}, which asserts one can split the dependency and distribution apart without losing any information. Fig.~\ref{GNPR_projection} illustrates this projection for $N = 2$.

We leverage the propounded representation to build a suitable yet simple distance between random variables which is invariant under diffeomorphism.

\vspace{0.2cm}
\textbf{Distance $d_\theta$ between two random variables} Let $\theta \in [0,1]$. Let $(X,Y) \in \mathcal{V}^2$. Let $G = (G_{X},G_{Y})$, where $G_X$ and $G_Y$ are respectively $X$ and $Y$ marginal cdfs. We define the following distance
\begin{eqnarray}\label{distance_theta}
d_{\theta}^{2}(X,Y)=\theta d_{1}^{2}(G_{X}(X),G_{Y}(Y))+(1-\theta)d_{0}^{2}(G_{X},G_{Y}),
\end{eqnarray}
where
\begin{eqnarray}
d_{1}^{2}(G_X(X),G_Y(Y))=3\mathbb{E}[\vert G_{X}(X)-G_{Y}(Y) \vert^{2}],
\end{eqnarray}
and
\begin{eqnarray}
d_{0}^{2}(G_{X},G_{Y})=\frac{1}{2}\int_{\mathbf{R}} \left(\sqrt{\frac{dG_{X}}{d\lambda}}-\sqrt{\frac{dG_{Y}}{d\lambda}}\right)^2 \, \mathrm{d}\lambda.
\end{eqnarray}

In particular, $d_0 = \sqrt{1 - BC}$ is the Hellinger distance related to the Bhattacharyya (1/2-Chernoff) coefficient $BC$ upper bounding the Bayes' classification error.
To quantify distribution dissimilarity, $d_0$ is used rather than the more general $\alpha$-Chernoff divergences since it satisfies the invariance to a monotonous transform of the variables (significant for practitioners as it ensures to be insensitive to scaling (e.g. choice of units) or measurement scheme (e.g. device, mathematical modelling) of the underlying phenomenon). In addition, $d_\theta$ can thus be efficiently implemented as a scalar product.
$d_1 = \sqrt{(1 - \rho_S) / 2}$ is a distance correlation measuring statistical dependence between two random variables, where $\rho_S$ is the Spearman's correlation between $X$ and $Y$. Notice that $d_1$ can be expressed by using the copula $C : [0,1]^2 \rightarrow [0,1]$ implicitly defined by the relation $G(X,Y) = C(G_X(X),G_Y(Y))$ since $\rho_S(X,Y) = 12 \int_{0}^{1}\int_{0}^{1} C(u,v) ~\mathrm{d}u ~\mathrm{d}v - 3$ \cite{fredricks2007relationship}.
\begin{exmp}[Distance $d_\theta$ between two Gaussians]
%Let $X \sim \mathcal{N}(\mu_X,\sigma_X^2)$ and $Y \sim \mathcal{N}(\mu_Y,\sigma_Y^2)$, with correlation $\rho(X,Y) = \rho$. Let $C$ be the Gaussian copula of $(X,Y)$.
Let $(X,Y)$ be a bivariate Gaussian vector, with $X \sim \mathcal{N}(\mu_X,\sigma_X^2)$, $Y \sim \mathcal{N}(\mu_Y,\sigma_Y^2)$ and $\rho(X,Y) = \rho$.
We obtain, 
$$d_\theta^2(X,Y) = \theta \frac{1 - \rho_S}{2} + (1-\theta) \left(1 - \sqrt{\frac{2 \sigma_X \sigma_Y}{\sigma_X^2 + \sigma_Y^2}}e^{-\frac{1}{4} \frac{(\mu_X - \mu_Y)^2}{\sigma_X^2 + \sigma_Y^2}}\right).$$
%Depending on the dependence structure between $X$ and $Y$, one can derive analytic expression for $\rho_S$ from the copula $C$ \cite{embrechts2003modelling}. Observe that $\rho_S = \frac{6}{\pi}\arcsin(\frac{\rho}{2})$ for both the Gaussian and Student-t copula \cite{aas2004modelling}.
\end{exmp}
Remember that for perfectly correlated Gaussians ($\rho = \rho_S = 1$), we want to discriminate on their distributions. We can observe that
\begin{itemize}
\item for $\sigma_X,\sigma_Y \rightarrow +\infty$, then $d_0(X,Y) \rightarrow 0$, it alleviates a main shortcoming of the basic $L_2$ distance which is diverging to $+\infty$ in this case;
\item if $\mu_X \neq \mu_Y$, for $\sigma_X,\sigma_Y \rightarrow 0$, then $d_0(X,Y) \rightarrow 1$, its maximum value, i.e. it means that two Gaussians cannot be more remote from each other than two different Dirac delta functions.
\end{itemize}
This distance is a fast and good proxy for distance $d_\theta$ when the first two moments $\mu$ and $\sigma$ predominate. Nonetheless, for datasets which contain heavy-tailed distributions, it fails to capture this information.

To apply the propounded distance $d_\theta$ on sampled data without parametric assumptions, we have to define its statistical estimate $\tilde{d}_\theta$ working on realizations of the i.i.d. random variables. Distance $d_1$ working with continuous uniform distributions can be approximated by normalized rank statistics yielding to discrete uniform distributions, in fact coordinates of the multivariate empirical copula \cite{deheuvels1979fonction} which is a non-parametric estimate converging uniformly toward the underlying copula \cite{deheuvels1981asymptotic}.
Distance $d_0$ working with densities can be approximated by using its discrete form working on histogram density estimates.

\textbf{The empirical copula transform.}
Let $X^T = (X_1^t,\ldots,X_N^t)$, $t=1,\ldots,T$, be $T$ observations from a random vector $X = (X_1,\ldots,X_N)$ with continuous margins $G_X = (G_{X_1}(X_1),\ldots,G_{X_N}(X_N))$. Since one cannot directly obtain the corresponding copula observations $(G_{X_1}(X_1^t),\ldots,G_{X_N}(X_N^t))$ without knowing a priori $G_X$, one can instead estimate the $N$ empirical margins $G_{X_i}^T(x) = \frac{1}{T}\sum_{t=1}^T \mathbf{1}(X_i^t \leq x)$ to obtain $T$ empirical observations $(G_{X_1}^T(X_1^t),\ldots,G_{X_N}^T(X_N^t))$ which are thus related to normalized rank statistics as $G_{X_i}^T(X_i^t) = X_i^{(t)} / T$, where $X_i^{(t)}$ denotes the rank of observation $X_i^t$.

%\begin{rmk}[Bijective ranking function]
%Let $(X_i)_{i=1}^M$ be $M$ realizations of $X \in \mathcal{V}$.
%Let $\mathfrak{S}_M$ be the permutation group of $\{1,\ldots,M\}$ and let $\sigma \in \mathfrak{S}_M$ be any fixed permutation, say $\sigma = Id_{\{1,\ldots,M\}}$. A bijective ranking function for $(X_i)_{i=1}^M$ can be defined as a function
%\begin{eqnarray}
%\mathrm{rk}^X : \{1,\ldots,M\} & \rightarrow & \{1,\ldots,M\} \\
%i & \mapsto & \# \{ k \in \{1,\ldots,M\} ~|~ \mathcal{P}_\sigma \} \nonumber
%\end{eqnarray}
%where $\mathcal{P}_\sigma \equiv (X_k < X_i) \lor (X_k = X_i \land \sigma(k) \leq \sigma(i))$.
%\end{rmk}
\textbf{Empirical distance.}
Let $(X^t)_{t=1}^T$ and $(Y^t)_{t=1}^T$ be $T$ realizations of real-valued random variables $X, Y \in \mathcal{V}$ respectively. 
An empirical distance between realizations of random variables can be defined by
\begin{eqnarray}
\tilde{d}_\theta^2\left((X^t)_{t=1}^T,(Y^t)_{t=1}^T\right) \stackrel{a.s.}{=} \theta \tilde{d}_{1}^2 + (1-\theta) \tilde{d}_{0}^2,
\end{eqnarray}
where
\begin{eqnarray}
\tilde{d}_1^2 = \frac{3}{T(T^2-1)}\sum_{t=1}^T \left(X^{(t)} - Y^{(t)}\right)^2
\end{eqnarray}
and
\begin{eqnarray}
\tilde{d}_0^2=\frac{1}{2}\sum_{k=-\infty}^{+\infty} \left(\sqrt{g_X^h(hk)} - \sqrt{g_Y^h(hk)} \right)^2,
\end{eqnarray}
$h$ being here a suitable bandwidth, and $g_X^h(x) = \frac{1}{T}\sum_{t=1}^T\mathbf{1}( \lfloor \frac{x}{h} \rfloor h\leq X^t < (\lfloor \frac{x}{h} \rfloor +1)h )$ being a density histogram estimating pdf $g_X$ from $(X^t)_{t=1}^T$, $T$ realizations of random variable $X \in \mathcal{V}$.

%We will refer henceforth to this distance and its use as the generic non-parametric representation (GNPR) approach. 
To use effectively $d_{\theta}$ and its statistical estimate, it boils down to select a particular value for $\theta$. We suggest here an exploratory approach where one can test (i) distribution information ($\theta = 0$), (ii) dependence information ($\theta = 1$), and (iii) a mix of both information ($\theta = 0.5$).
Ideally, $\theta$ should reflect the balance of dependence and distribution information in the data. 
In a supervised setting, one could select an estimate $\hat{\theta}$ of the right balance $\theta^\star$ optimizing some loss function by techniques such as cross-validation.
Yet, the lack of a clear loss function makes the estimation of $\theta^\star$ difficult in an unsupervised setting. For clustering, many authors \cite{lange2004stability}, \cite{shamir2007cluster}, \cite{shamir2008model}, \cite{meinshausen2010stability} suggest stability as a tool for parameter selection. 

%But, authors in \cite{ben2006sober} warn against its irrelevant use for this purpose. Besides, we already use stability for clustering validation \cite{marti2015proposal} and we want to avoid overfitting.
%Finally, we think that finding an optimal trade-off $\theta^\star$ is important for accelerating the rate of convergence toward the underlying ground truth (for example, a hierarchical block model with correlation blocks subdivided into distribution blocks) when working with finite and possibly small samples, but ultimately lose its importance asymptotically as soon as $0 < \theta < 1$.

\subsection{How to extend the approach to $N$ multivariate time series?}

We are now interested in clustering $N$ assets which are described by more than one time series.
Though a stock is usually described by a single time series, its market price, other assets such as credit default swaps can be described by several maturities, their term structure. In practice, a CDS term structure time series is a $5$-variate time series. At each time $t$, it consists in $d=5$ prices for the different traded maturities: $1, 3, 5, 7, 10$ years. 
In our opinion, the case where each object is described by several time series has not been thoroughly explored in the machine learning literature \cite{yang2004pca,singhal2002clustering,dasu2005grouping}. We suggest ways to develop a geometry based methodology to address this clustering problem.
At least three avenues of research can be explored: 
\begin{itemize}
\item distances from Information Geometry theory,
\item distances from Optimal Transport theory,
\item distances from kernel embedding of distributions \cite{smola2007hilbert}.
\end{itemize}

\subsubsection{Intra-dependence and margins.}

We suppose that the $d$ time series describing a given asset follow a $d$-variate distribution of density $f(x) := f(x_1,\ldots,x_d)$.
According to Sklar's Theorem \cite{sklar1959fonctions}, we have 
\begin{equation}
f(x_1,\ldots,x_d) = c(F_1(x_1),\ldots,F_d(x_d))\prod_{i=1}^d f_i(x_i),
\end{equation}
where $c$ is the copula density, $F_i$ are the marginal cumulative distribution functions and $f_i$ their densities.

Assuming a parametric modelling, we can derive the Fisher-Rao geodesic distance between two assets represented by their parametric multivariate densities $f(x_1,\ldots,x_d;\theta_1)$ and $f(x_1,\ldots,x_d;\theta_2)$ respectively. Since the copula density $c$ has its own set of parameters $\theta_c$ and the margins $f_i$ also have their own parameters $\theta_{m_i}$, we have $f(x_1,\ldots,x_d;\theta) = f(x_1,\ldots,x_d;\theta_c,\theta_m)$ which is equal to $c(F_1(x_1;\theta_{m_1}),\ldots,F_d(x_d;\theta_{m_d});\theta_c) \prod_{i=1}^d f_i(x_i;\theta_{m_i})$. To compute the Fisher-Rao geodesic distance $D$ between $f(x;\theta_1)$ and $f(x;\theta_2)$:
\begin{equation}D(f(x;\theta_1),f(x;\theta_2)) =  \int_{\theta_1}^{\theta_2} ds = \int_{0}^{1} \sqrt{ \sum_{i,j} g_{ij}(\theta(t)) \frac{d\theta^{i}}{dt} \frac{d\theta^{j}}{dt} } dt,
\end{equation}
we first compute the Fisher information matrix $g_{ij}(\theta)$:

\begin{eqnarray}
g_{ij}(\theta) & = & - \mathbb{E}_X\left[ \frac{\partial^2}{\partial \theta^i \partial \theta^j} \log c(F_1(x_1;\theta_{m_1}),\ldots,F_d(x_d;\theta_{m_d});\theta_{c}) \right] \\
&  & - \mathbb{E}_X\left[ \frac{\partial^2}{\partial \theta^i \partial \theta^j} \log \prod_{k=1}^d f_k(x_k;\theta_{m_k}) \right] \\
& = & - \mathbb{E}_X\left[ \frac{\partial^2}{\partial \theta^i \partial \theta^j} \log c(F_1(x_1;\theta_{m_1}),\ldots,F_d(x_d;\theta_{m_d});\theta_{c}) \right] \\
&  & - \sum_{k=1}^{d} \mathbb{E}_X\left[ \frac{\partial^2}{\partial \theta^i \partial \theta^j} \log f_k(x_k;\theta_{m_k}) \right]
\end{eqnarray}

%$$= - \mathbb{E}_X\left[ \frac{\partial^2}{\partial \theta^i \partial \theta^j} \log c(F_1(x_1;\theta_{m_1}),\ldots,F_d(x_d;\theta_{m_d});\theta_{c}) \right] - \mathbb{E}_X\left[ \frac{\partial^2}{\partial \theta^i \partial \theta^j} \log \prod_{i=1}^d f_i(x_i;\theta_{m_i}) \right]$$

%$$ = - \mathbb{E}_X\left[ \frac{\partial^2}{\partial \theta^i \partial \theta^j} \log c(F_1(x_1;\theta_{m_1}),\ldots,F_d(x_d;\theta_{m_d});\theta_{c}) \right] - \sum_{i=1}^{d} \mathbb{E}_X\left[ \frac{\partial^2}{\partial \theta^i \partial \theta^j} \log f_i(x_i;\theta_{m_i}) \right]$$

If we opt for the Canonical Maximum Likelihood hypothesis as in \cite{el2011color}, then $\frac{\partial}{\partial \theta_m} c(u_1,\ldots,u_d;\theta_c) = 0$. It follows that $g_{\theta_c, \theta_m} = g_{\theta_m, \theta_c} = 0$. Thus, we obtain the Fisher-Rao metric 
\begin{eqnarray}
ds^2 = \sum_{i,j} g_{ij}(\theta) d\theta^i d\theta^j = g_{\theta_c,\theta_c}d\theta_c d\theta_c + \sum_{i=1}^d \sum_{k,l} g_{\theta_{m_k},\theta_{m_l}} d\theta_{m_l} d\theta_{m_k}.
\end{eqnarray}
It can be expressed by 
\begin{equation}
ds^2 = ds^2_{copula} + \sum_{i=1}^d ds^2_{margins},
\end{equation}
and therefore the Fisher-Rao geodesic distance is a distance between the dependence structure of the two multivariate densities + a distance between the marginal distributions of these two multivariate densities.

However, since the Fisher-Rao distance is frequently intractable, one often considers related divergences such as Kullback-Leibler, symmetrized Jeffreys, Hellinger, or Bhattacharyya divergences which coincide with the quadratic form approximations of the Fisher-Rao distance between two close distributions, and which are computationally more tractable. It would be interesting to find the class of divergences that verifies such a decomposability. For instance, the Kullback-Leibler divergence does not: $KL(f,g) \neq KL(c_f,c_g) + \sum_{i=1}^d KL(f_i,g_i)$.
However, if $f$ and $g$ have identical marginals, i.e. $\forall i \in \{1,\ldots,d\}, f_i = g_i$, then it can be shown \cite{killiches2015model} that $KL(f,g) = KL(c_f,c_g) = KL(c_f,c_g) + \sum_{i=1}^d KL(f_i,g_i)$.

\begin{figure}
\begin{center}
\includegraphics[width=0.32\linewidth]{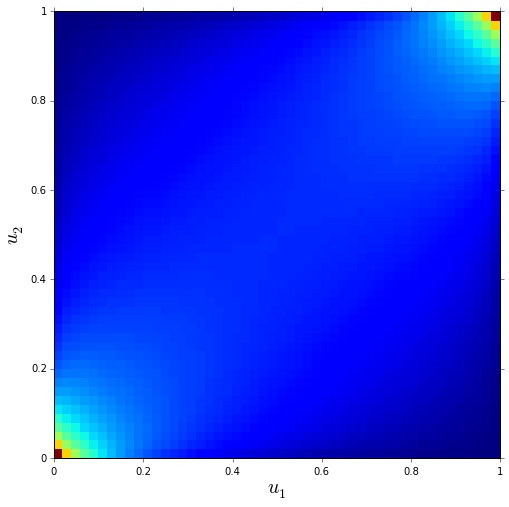}
\includegraphics[width=0.32\linewidth]{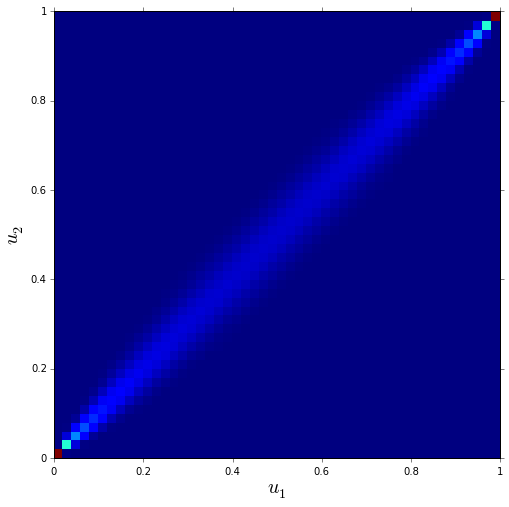}
\includegraphics[width=0.32\linewidth]{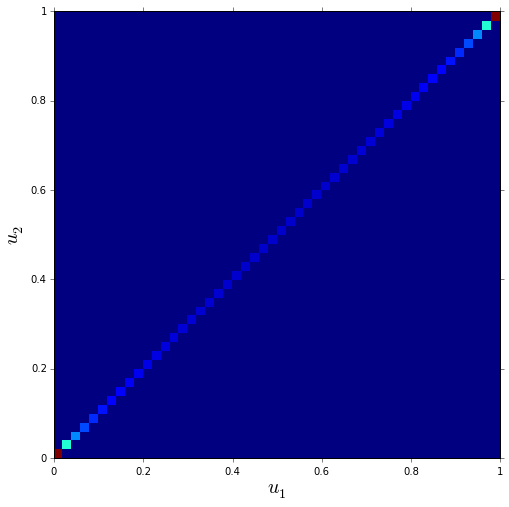}
\end{center}
\caption{Densities of $C_{R_A}^{\mathrm{Gauss}}, C_{R_B}^{\mathrm{Gauss}}, C_{R_C}^{\mathrm{Gauss}}$ respectively;  Notice that for strong correlations, the density tends to be distributed very close to the diagonal.}
\label{fig:copula_pdf}
\end{figure}

How the choice of a particular distance will influence the clustering? A brief comparison of Fisher-Rao and its related divergences and the Wasserstein $W_2$ distance between bivariate Gaussian copulas is provided for illustration. Let $C_{R_A}^{\mathrm{Gauss}}$, $C_{R_B}^{\mathrm{Gauss}}$, $C_{R_C}^{\mathrm{Gauss}}$ be three bivariate Gaussian copulas parameterized by the following correlation matrices
 $$R_A = \begin{pmatrix}
   1 & 0.5 \\
   0.5 & 1 
\end{pmatrix}, R_B = \begin{pmatrix}
   1 & 0.99 \\
   0.99 & 1 
\end{pmatrix}, R_C = \begin{pmatrix}
   1 & 0.9999 \\
   0.9999 & 1 
\end{pmatrix}$$ respectively. Heatmaps of their densities are plotted in Fig.~\ref{fig:copula_pdf}.

In Table~\ref{tab:distances}, we report the distances $D(R_A,R_B)$ between $C_{R_A}^{\mathrm{Gauss}}$ and $C_{R_B}^{\mathrm{Gauss}}$, and the distances $D(R_B,R_C)$ between $C_{R_B}^{\mathrm{Gauss}}$ and $C_{R_C}^{\mathrm{Gauss}}$.
We can observe that unlike Wasserstein $W_2$ distance, Fisher-Rao and related divergences consider that $C_{R_A}^{\mathrm{Gauss}}$ and $C_{R_B}^{\mathrm{Gauss}}$ are nearer than $C_{R_B}^{\mathrm{Gauss}}$ and $C_{R_C}^{\mathrm{Gauss}}$.
This may be an undesirable property for clustering since $C_{R_B}^{\mathrm{Gauss}}$ and $C_{R_C}^{\mathrm{Gauss}}$ both describe a strong positive dependence between the two variates whereas $C_{R_A}^{\mathrm{Gauss}}$ describes only a mild positive dependence.

\begin{table*}
\caption{Distances in closed-form between Gaussians and their sensitivity to the correlation strength}
\begin{center}
\begin{tabular}{ccccc}
    \toprule
           & $D\left(\mathcal{N}(0,\Sigma_1),\mathcal{N}(0,\Sigma_2)\right)$ & $D(R_A,R_B)$ & & $D(R_B,R_C)$ \\ \midrule
    Fisher-Rao \cite{atkinson1981rao} & $\sqrt{ \frac{1}{2}\sum_{i=1}^n (\log \lambda_i)^2 }$  & 2.77 & $<$ & 3.26 \\
    $KL(\Sigma_1||\Sigma_2)$ & $\frac{1}{2} \left( \log \frac{|\Sigma_2|}{|\Sigma_1|} - n + tr(\Sigma_2^{-1} \Sigma_1) \right) $ & 22.6 & $<$ & 47.2 \\
    Jeffreys & $KL(\Sigma_1||\Sigma_2) + KL(\Sigma_2||\Sigma_1)$  & 24 & $<$ & 49    \\
    Hellinger & $\sqrt{ 1 - \frac{|\Sigma_1|^{1/4} |\Sigma_2|^{1/4}}{|\Sigma|^{1/2}} }$ & 0.48 & $<$ & 0.56    \\
    Bhattacharyya & $\frac{1}{2} \log \frac{|\Sigma|}{\sqrt{|\Sigma_1| |\Sigma_2| } } $   & 0.65 & $<$ & 0.81    \\
    $W_2$ \cite{takatsu2011wasserstein} & $\sqrt{ tr\left( \Sigma_1 + \Sigma_2 - 2\sqrt{\Sigma_1^{1/2} \Sigma_2 \Sigma_1^{1/2}} \right) }$ & \textbf{0.63} & $\mathbf{>}$ & \textbf{0.09}   \\ \bottomrule
  \end{tabular}
  
  $\lambda_i$ eigenvalues of $\Sigma_1^{-1}\Sigma_2$; $\Sigma = \frac{\Sigma_1 + \Sigma_2}{2}$
\end{center}\label{tab:distances}
\end{table*}

\begin{figure}
\begin{center}
\includegraphics[width=0.15\linewidth]{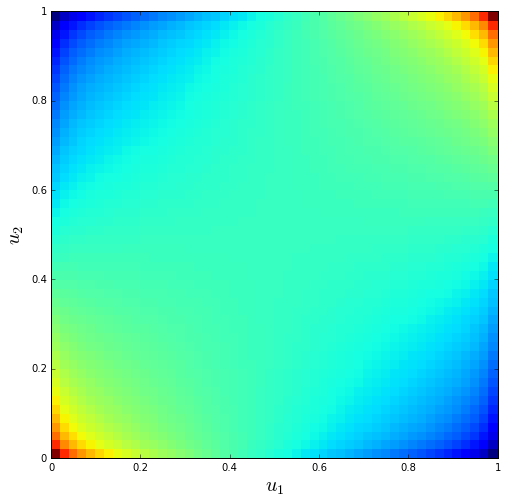}
\includegraphics[width=0.15\linewidth]{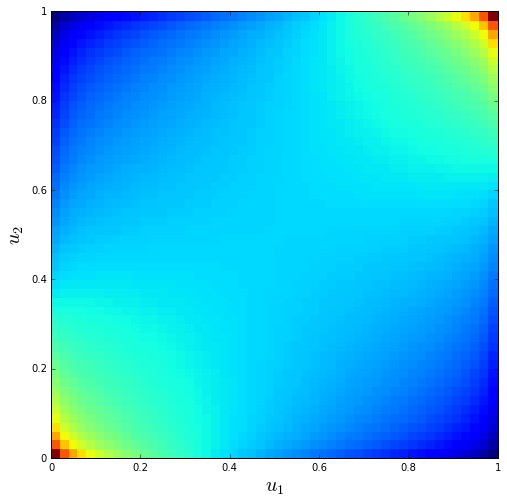}
\includegraphics[width=0.15\linewidth]{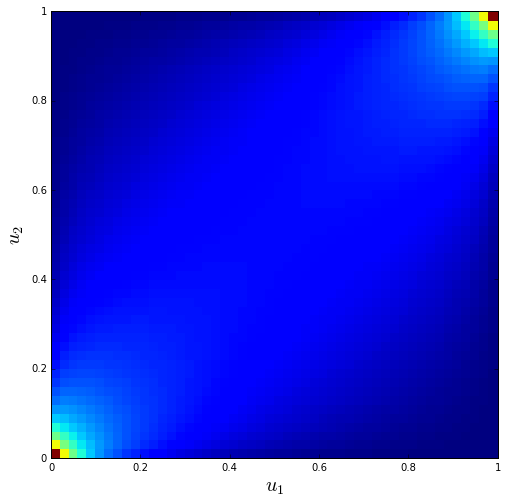}
\includegraphics[width=0.15\linewidth]{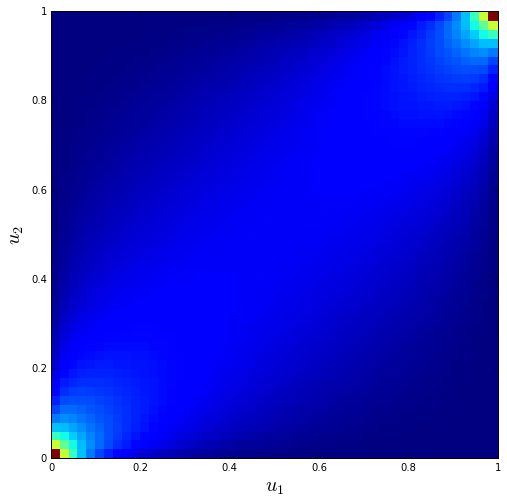}
\includegraphics[width=0.15\linewidth]{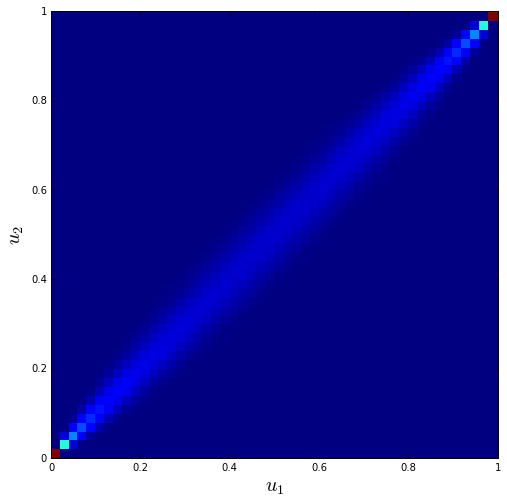}
\includegraphics[width=0.15\linewidth]{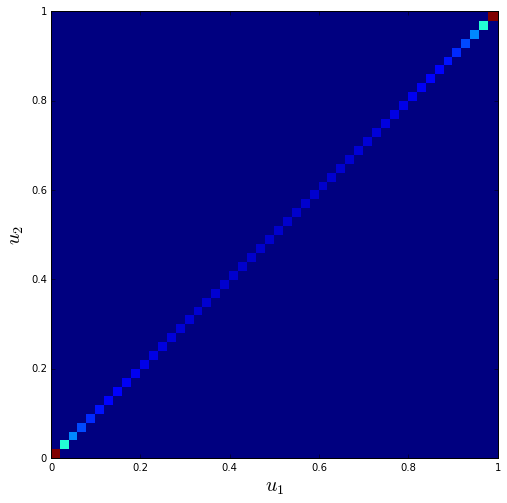}
\end{center}
\caption{Datasets of bivariate time series are generated from six Gaussian copulas with correlation .1, .2, .6, .7, .99, .9999}\label{fig:correl_clusters}
\end{figure}

\begin{figure}
\begin{center}
\includegraphics[width=0.46\linewidth]{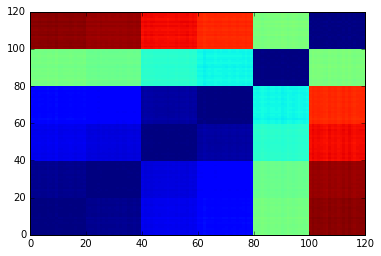}
\includegraphics[width=0.46\linewidth]{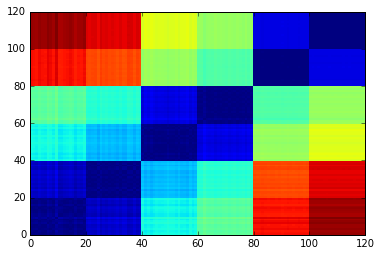}
\end{center}
\caption{Distance heatmaps for Fisher-Rao (left), $W_2$ (right); Using Ward clustering, Fisher-Rao yields clusters of copulas with correlations $\{.1,.2,.6,.7\}$, $\{.99\}$, $\{.9999\}$, $W_2$ yields $\{.1,.2 \}$, $\{.6,.7 \}$, $\{.99,.9999 \}$ }\label{fig:dist_clusters}
\end{figure}

In financial applications, variates can be strongly correlated (for instance, the returns of different maturities in a term structure).
In such cases, Fisher-Rao and related divergences yield a much different clustering than the one obtained from using a Wasserstein $W_2$ distance: Let's consider a dataset of $N$ bivariate time series evenly generated from the six Gaussian copulas depicted in Fig.~\ref{fig:correl_clusters}.
When a clustering algorithm such as Ward is given a distance matrix computed from Fisher-Rao (displayed in Fig.~\ref{fig:dist_clusters}), it will tend to gather in a cluster all copulas but the ones describing high dependence which are isolated. $W_2$ yields a more balanced and intuitive clustering where clusters contain copulas of similar dependence.

Thus, if the dependence is strong between the time series, the use of Fisher-Rao geodesic distance and related divergences may not be appropriate. 
They are relevant to find which samples were generated from the same set of parameters (clustering viewed as a generalization of the three-sample problem \cite{ryabko2010clustering}) due to their local expression as a quadratic form of the Fisher Information Matrix determining the Cram\'er-Rao Lower Bound on the variance of estimators. To measure distance between copulas for clustering purpose, Wasserstein geometry may be more appropriate since it does not lead to these counter-intuitive clusters.
We will investigate further this issue. We would also like to encompass the embedding of probability distributions into reproducing kernel Hilbert spaces \cite{sriperumbudur2009kernel} in our comparison of the possible distances for copulas.

\subsubsection{Inter-dependence.} 
\begin{figure}[htp]
\begin{center}
\includegraphics[width=\linewidth]{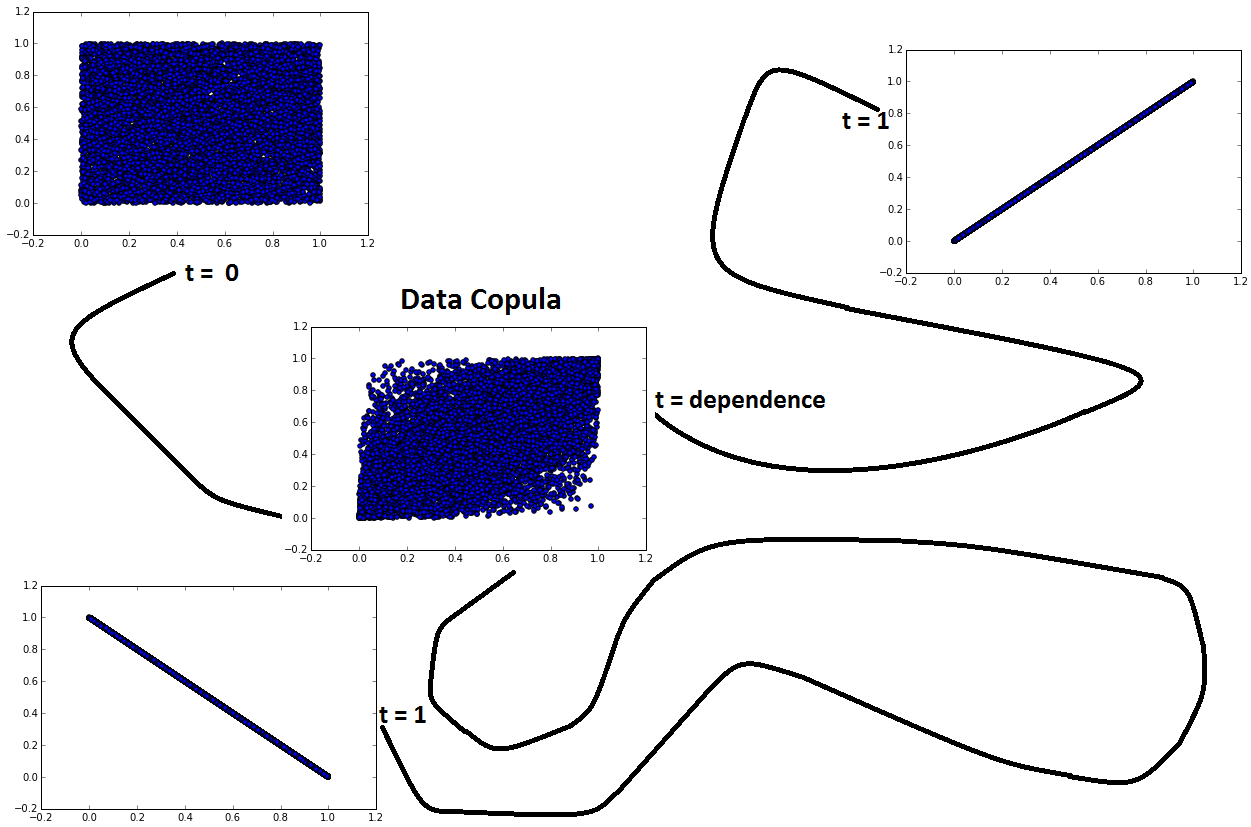}
\caption{Dependence can be seen as the relative distance between the independence copula and one or more target dependence copulas. In this picture, the target dependencies are ``perfect dependence" and ``perfect anti-dependence". The empirical copula (Data Copula) was built from positively correlated Gaussians, and thus is nearer to the ``perfect dependence" copula (top right corner) than to the ``perfect anti-dependence" copula (bottom left corner).}\label{fig:copula_transport}
\end{center}
\end{figure}

However, notice that the distance between the two copulas only measures the difference in the coordinates $x_1,\ldots,x_d$ joint behaviour of their respective multivariate distribution, i.e. the intra-dependence. It gives no information on the time series joint behaviour (how are they moving together?)
To obtain such information, one could build the $2d$-variate copula of the two $d$-variate time series viewed as a single $2d$-variate time series and compare it to the $2d$-variate independence copula (this idea is depicted in Fig.~\ref{fig:copula_transport}). Such an approach, using optimal transport to compare copulas, is described in \cite{marti2015optimal}.
But, this construction captures a mixed information of intra-dependence (the coordinates joint behaviour) and inter-dependence (the multivariate time series joint behaviour), besides losing the notion of two different time series. It has been shown that copula is an inadequate tool to build distributions with multivariate marginals \cite{genest1995impossibilite}.
In \cite{li1996linkages}, authors propose an analogous tool called the linkage function to address these problems: the linkage function contains the information regarding the dependence structure among the underlying multivariate distributions (inter-dependence) but the dependence structure within the multivariate distributions (intra-dependence) is not included.

\section{Discussion}

In this work, we have presented a new modelling framework for studying financial time series.
Clustering could allow to develop an alternative portfolio theory and more relevant risk measures.
Several researchers have begun to explore this avenue of research.
Until now they have used the Pearson correlation matrix as a similarity matrix for clustering the assets, and thus assuming the Gaussianity of the log-returns. % (the geometric brownian motion assumption used in the Black-Scholes model implies that the log-returns are normally distributed). 
We propose to replace the Pearson correlation matrix by a matrix whose coefficients measure more accurately the dependence and distributional similarities between the assets' returns which can follow any arbitrary joint distribution.
For the Information Geometry theoretician, it boils down to design distances between dependent random variables.
We think that an interesting approach could be achieved by developing a geometry based on the (copula, margins) representation for random variables, and maybe a (linkage, (copula, margins)) representation for random vectors.
We have already started to experiment with (regularized) optimal transport and look forward to leverage information geometry distances to improve our clustering methodology of financial time series. We will be glad to obtain more feedback and hope that our problem was exposed clearly enough so other researchers can work on developing a proper geometry for these dependent (multivariate) distributions.

\subsubsection*{Acknowledgments.} Gautier Marti wants to thank Prof. Eguchi for helpful and encouraging remarks, Prof. Brigo for pointing us interesting research directions on dependence, copulas and optimal transport, and Fr\'ed\'eric Barbaresco for sending us relevant literature, historical references, and interesting discussions.
We also want to thank our colleagues at Hellebore Capital, and the friendly feedbacks from Philippe Very. Finally, the authors thank the organizers of the workshop ``Computational information geometry for image and signal processing" at the International Centre for Mathematical Sciences, Edinburgh, UK, for the invitation.

\bibliographystyle{plain}
\bibliography{biblio}

\begin{thebibliography}{10}

\bibitem{allez2014eigenvectors}
Romain Allez, Jo{\"e}l Bun, and Jean-Philippe Bouchaud.
\newblock The eigenvectors of gaussian matrices with an external source.
\newblock {\em arXiv preprint arXiv:1412.7108}, 2014.

\bibitem{ao2005clustag}
Sio~Iong Ao, Kevin Yip, Michael Ng, David Cheung, Pui-Yee Fong, Ian Melhado,
  and Pak~C Sham.
\newblock Clustag: hierarchical clustering and graph methods for selecting tag
  snps.
\newblock {\em Bioinformatics}, 21(8):1735--1736, 2005.

\bibitem{atkinson1981rao}
Colin Atkinson and Ann~FS Mitchell.
\newblock Rao's distance measure.
\newblock {\em Sankhy{\=a}: The Indian Journal of Statistics, Series A}, pages
  345--365, 1981.

\bibitem{balakrishnan2011noise}
Sivaraman Balakrishnan, Min Xu, Akshay Krishnamurthy, and Aarti Singh.
\newblock Noise thresholds for spectral clustering.
\newblock pages 954--962, 2011.

\bibitem{basalto2007hausdorff}
Nicolas Basalto, Roberto Bellotti, Francesco De~Carlo, Paolo Facchi, Ester
  Pantaleo, and Saverio Pascazio.
\newblock Hausdorff clustering of financial time series.
\newblock {\em Physica A: Statistical Mechanics and its Applications},
  379(2):635--644, 2007.

\bibitem{bien2011hierarchical}
Jacob Bien and Robert Tibshirani.
\newblock Hierarchical clustering with prototypes via minimax linkage.
\newblock {\em Journal of the American Statistical Association},
  106(495):1075--1084, 2011.

\bibitem{borysov2014asymptotics}
Petro Borysov, Jan Hannig, and JS~Marron.
\newblock Asymptotics of hierarchical clustering for growing dimension.
\newblock {\em Journal of Multivariate Analysis}, 124:465--479, 2014.

\bibitem{bun2015rotational}
Jo{\"e}l Bun, Romain Allez, Jean-Philippe Bouchaud, and Marc Potters.
\newblock Rotational invariant estimator for general noisy matrices.
\newblock {\em arXiv preprint arXiv:1502.06736}, 2015.

\bibitem{chen1996space}
Zhenmin Chen and John~W Van~Ness.
\newblock Space-conserving agglomerative algorithms.
\newblock {\em Journal of classification}, 13(1):157--168, 1996.

\bibitem{cont2001empirical}
Rama Cont.
\newblock Empirical properties of asset returns: stylized facts and statistical
  issues.
\newblock 2001.

\bibitem{costa2014fisher}
Sueli~IR Costa, Sandra~A Santos, and Jo{\~a}o~E Strapasson.
\newblock Fisher information distance: a geometrical reading.
\newblock {\em Discrete Applied Mathematics}, 2014.

\bibitem{dasu2005grouping}
Tamraparni Dasu, Deborah~F Swayne, and David Poole.
\newblock Grouping multivariate time series: A case study.
\newblock In {\em Proceedings of the IEEE Workshop on Temporal Data Mining:
  Algorithms, Theory and Applications, in conjunction with the Conference on
  Data Mining, Houston}, pages 25--32, 2005.

\bibitem{deheuvels1979fonction}
Paul Deheuvels.
\newblock La fonction de d{\'e}pendance empirique et ses propri{\'e}t{\'e}s.
  {U}n test non param{\'e}trique d{'}ind{\'e}pendance.
\newblock {\em Acad. Roy. Belg. Bull. Cl. Sci.(5)}, 65(6):274--292, 1979.

\bibitem{deheuvels1981asymptotic}
Paul Deheuvels.
\newblock An asymptotic decomposition for multivariate distribution-free tests
  of independence.
\newblock {\em Journal of Multivariate Analysis}, 11(1):102--113, 1981.

\bibitem{donnat2016toward}
Philippe Donnat, Gautier Marti, and Philippe Very.
\newblock Toward a generic representation of random variables for machine
  learning.
\newblock {\em Pattern Recognition Letters}, 70:24--31, 2016.

\bibitem{el2011color}
Ahmed~Drissi El~Maliani, Mohammed El~Hassouni, Nour-Eddine Lasmar, Yannick
  Berthoumieu, and Driss Aboutajdine.
\newblock Color texture classification using rao distance between multivariate
  copula based models.
\newblock In {\em Computer Analysis of Images and Patterns}, pages 498--505.
  Springer, 2011.

\bibitem{fredricks2007relationship}
Gregory~A Fredricks and Roger~B Nelsen.
\newblock On the relationship between spearman's rho and kendall's tau for
  pairs of continuous random variables.
\newblock {\em Journal of Statistical Planning and Inference},
  137(7):2143--2150, 2007.

\bibitem{genest1995impossibilite}
Christian Genest, JJ~Quesada~Molina, and JA~Rodr{\'\i}guez~Lallena.
\newblock De l'impossibilit{\'e} de construire des lois {\`a} marges
  multidimensionnelles donn{\'e}es {\`a} partir de copules.
\newblock {\em Comptes rendus de l'Acad{\'e}mie des sciences. S{\'e}rie 1,
  Math{\'e}matique}, 320(6):723--726, 1995.

\bibitem{hartigan1981consistency}
John~A Hartigan.
\newblock Consistency of single linkage for high-density clusters.
\newblock {\em Journal of the American Statistical Association},
  76(374):388--394, 1981.

\bibitem{khaleghi2012online}
Azadeh Khaleghi, Daniil Ryabko, Jeremie Mary, and Philippe Preux.
\newblock Online clustering of processes.
\newblock pages 601--609, 2012.

\bibitem{khaleghi2016consistent}
Azadeh Khaleghi, Daniil Ryabko, J{\'e}r{\'e}mie Mary, and Philippe Preux.
\newblock Consistent algorithms for clustering time series.
\newblock {\em Journal of Machine Learning Research}, 17(3):1--32, 2016.

\bibitem{killiches2015model}
Matthias Killiches, Daniel Kraus, and Claudia Czado.
\newblock Model distances for vine copulas in high dimensions with application
  to testing the simplifying assumption.
\newblock {\em arXiv preprint arXiv:1510.03671}, 2015.

\bibitem{krishnamurthy2012efficient}
Akshay Krishnamurthy, Sivaraman Balakrishnan, Min Xu, and Aarti Singh.
\newblock Efficient active algorithms for hierarchical clustering.
\newblock {\em International Conference on Machine Learning}, 2012.

\bibitem{laloux1999noise}
Laurent Laloux, Pierre Cizeau, Jean-Philippe Bouchaud, and Marc Potters.
\newblock Noise dressing of financial correlation matrices.
\newblock {\em Physical review letters}, 83(7):1467, 1999.

\bibitem{laloux2000random}
Laurent Laloux, Pierre Cizeau, Marc Potters, and Jean-Philippe Bouchaud.
\newblock Random matrix theory and financial correlations.
\newblock {\em International Journal of Theoretical and Applied Finance},
  3(03):391--397, 2000.

\bibitem{lange2004stability}
Tilman Lange, Volker Roth, Mikio~L Braun, and Joachim~M Buhmann.
\newblock Stability-based validation of clustering solutions.
\newblock {\em Neural computation}, 16(6):1299--1323, 2004.

\bibitem{lemieux2014clustering}
Victoria Lemieux, Payam~S Rahmdel, Rick Walker, BL~Wong, and Mark Flood.
\newblock Clustering techniques and their effect on portfolio formation and
  risk analysis.
\newblock pages 1--6, 2014.

\bibitem{li1996linkages}
Haijun Li, Marco Scarsini, and Moshe Shaked.
\newblock Linkages: a tool for the construction of multivariate distributions
  with given nonoverlapping multivariate marginals.
\newblock {\em Journal of Multivariate Analysis}, 56(1):20--41, 1996.

\bibitem{mantegna1999hierarchical}
Rosario~N Mantegna.
\newblock Hierarchical structure in financial markets.
\newblock {\em The European Physical Journal B-Condensed Matter and Complex
  Systems}, 11(1):193--197, 1999.

\bibitem{mantegna1999introduction}
Rosario~N Mantegna and H~Eugene Stanley.
\newblock {\em Introduction to econophysics: correlations and complexity in
  finance}.
\newblock Cambridge university press, 1999.

\bibitem{marti2016clustering}
Gautier Marti, S{\'e}bastien Andler, Frank Nielsen, and Philippe Donnat.
\newblock Clustering financial time series: How long is enough?
\newblock {\em arXiv preprint arXiv:1603.04017}, 2016.

\bibitem{marti2015optimal}
Gautier Marti, Frank Nielsen, and Philippe Donnat.
\newblock Optimal copula transport for clustering multivariate time series.
\newblock {\em IEEE ICASSP}, 2016.

\bibitem{marti2015proposal}
Gautier Marti, Philippe Very, Philippe Donnat, and Frank Nielsen.
\newblock A proposal of a methodological framework with experimental guidelines
  to investigate clustering stability on financial time series.
\newblock {\em IEEE ICMLA}, 2015.

\bibitem{meinshausen2010stability}
Nicolai Meinshausen and Peter B{\"u}hlmann.
\newblock Stability selection.
\newblock {\em Journal of the Royal Statistical Society: Series B (Statistical
  Methodology)}, 72(4):417--473, 2010.

\bibitem{murtagh2011methods}
Fionn Murtagh and Pedro Contreras.
\newblock Methods of hierarchical clustering.
\newblock {\em arXiv preprint arXiv:1105.0121}, 2011.

\bibitem{pantaleo2011improved}
Ester Pantaleo, Michele Tumminello, Fabrizio Lillo, and Rosario~N Mantegna.
\newblock When do improved covariance matrix estimators enhance portfolio
  optimization? an empirical comparative study of nine estimators.
\newblock {\em Quantitative Finance}, 11(7):1067--1080, 2011.

\bibitem{plerou2002random}
Vasiliki Plerou, Parameswaran Gopikrishnan, Bernd Rosenow, Luis A~Nunes Amaral,
  Thomas Guhr, and H~Eugene Stanley.
\newblock Random matrix approach to cross correlations in financial data.
\newblock {\em Physical Review E}, 65(6):066126, 2002.

\bibitem{pollard1981strong}
David Pollard et~al.
\newblock Strong consistency of $ k $-means clustering.
\newblock {\em The Annals of Statistics}, 9(1):135--140, 1981.

\bibitem{potters2005financial}
Marc Potters, Jean-Philippe Bouchaud, and Laurent Laloux.
\newblock Financial applications of random matrix theory: Old laces and new
  pieces.
\newblock {\em arXiv preprint physics/0507111}, 2005.

\bibitem{ryabko2010clustering}
D.~Ryabko.
\newblock Clustering processes.
\newblock {\em Proc. the 27th International Conference on Machine Learning
  (ICML 2010)}, pages 919--926, 2010.

\bibitem{shamir2007cluster}
Ohad Shamir and Naftali Tishby.
\newblock Cluster stability for finite samples.
\newblock In {\em NIPS}, 2007.

\bibitem{shamir2008model}
Ohad Shamir and Naftali Tishby.
\newblock Model selection and stability in k-means clustering.
\newblock In {\em Learning theory}, 2008.

\bibitem{singhal2002clustering}
Ashish Singhal and Dale~E Seborg.
\newblock Clustering of multivariate time-series data.
\newblock In {\em American Control Conference, 2002. Proceedings of the 2002},
  volume~5, pages 3931--3936. IEEE, 2002.

\bibitem{sklar1959fonctions}
A~Sklar.
\newblock {\em Fonctions de r{\'e}partition {\`a} n dimensions et leurs
  marges}.
\newblock Universit{\'e} Paris 8, 1959.

\bibitem{smola2007hilbert}
Alex Smola, Arthur Gretton, Le~Song, and Bernhard Sch{\"o}lkopf.
\newblock A hilbert space embedding for distributions.
\newblock In {\em Algorithmic Learning Theory}, pages 13--31. Springer, 2007.

\bibitem{sriperumbudur2009kernel}
Bharath~K Sriperumbudur, Kenji Fukumizu, Arthur Gretton, Gert~RG Lanckriet, and
  Bernhard Sch{\"o}lkopf.
\newblock Kernel choice and classifiability for rkhs embeddings of probability
  distributions.
\newblock In {\em NIPS}, pages 1750--1758, 2009.

\bibitem{takatsu2011wasserstein}
Asuka Takatsu et~al.
\newblock Wasserstein geometry of gaussian measures.
\newblock {\em Osaka Journal of Mathematics}, 48(4):1005--1026, 2011.

\bibitem{terada2013strong}
Yoshikazu Terada.
\newblock Strong consistency of factorial k-means clustering.
\newblock {\em Annals of the Institute of Statistical Mathematics},
  67(2):335--357, 2013.

\bibitem{terada2014strong}
Yoshikazu Terada.
\newblock Strong consistency of reduced k-means clustering.
\newblock {\em Scandinavian Journal of Statistics}, 41(4):913--931, 2014.

\bibitem{tola2008cluster}
Vincenzo Tola, Fabrizio Lillo, Mauro Gallegati, and Rosario~N Mantegna.
\newblock Cluster analysis for portfolio optimization.
\newblock {\em Journal of Economic Dynamics and Control}, 32(1):235--258, 2008.

\bibitem{tumminello2007shrinkage}
Michele Tumminello, Fabrizio Lillo, and Rosario~Nunzio Mantegna.
\newblock Shrinkage and spectral filtering of correlation matrices: a
  comparison via the kullback-leibler distance.
\newblock {\em arXiv preprint arXiv:0710.0576}, 2007.

\bibitem{von2008consistency}
Ulrike Von~Luxburg, Mikhail Belkin, and Olivier Bousquet.
\newblock Consistency of spectral clustering.
\newblock {\em The Annals of Statistics}, pages 555--586, 2008.

\bibitem{yang2004pca}
Kiyoung Yang and Cyrus Shahabi.
\newblock A {PCA}-based similarity measure for multivariate time series.
\newblock In {\em Proceedings of the 2nd ACM international workshop on
  Multimedia databases}, pages 65--74. ACM, 2004.

\end{thebibliography}

\end{document}